\documentclass{frontiersSCNS} 

\graphicspath{{./}{./figures/}}
\usepackage{url,lineno}

\copyrightyear{}
\pubyear{}

\def\journal{Robotics and AI}
\def\DOI{}
\def\articleType{Research Article}
\def\keyFont{\fontsize{8}{11}\helveticabold }
\def\firstAuthorLast{Echeveste {et~al.}} 
\def\Authors{Rodrigo Echeveste\,$^{1}$, Claudius Gros\,$^{1}$}
\def\Address{$^{1}$ Institute for Theoretical Physics, Goethe University Frankfurt, Germany.}
\def\corrAuthor{Claudius Gros}
\def\corrAddress{Goethe University Frankfurt, Germany}
\def\corrEmail{gros07@itp.uni-frankfurt.de}

\begin{document}
\onecolumn
\firstpage{1}

\title[The Fisher information as a generating functional]{Generating 
       functionals for computational intelligence:
       The Fisher information as an objective function for
       self-limiting Hebbian learning rules}
\author[\firstAuthorLast ]{\Authors}
\address{}
\correspondance{}
\extraAuth{}
\topic{}

\maketitle

\begin{abstract}
Generating functionals may guide the evolution of
a dynamical system and constitute a possible route 
for handling the complexity of neural networks as
relevant for computational intelligence. We propose and 
explore a new objective function which allows to
obtain plasticity rules for the afferent synaptic 
weights. The adaption rules are Hebbian, self-limiting,
and result from the minimization of the the Fisher 
information with respect to the synaptic flux.

We perform a series of simulations examining the behavior of 
the new learning rules in various circumstances. The vector 
of synaptic weights aligns with the principal direction of 
input activities, whenever one is present. A linear 
discrimination is performed when there are two or more principal 
directions; directions having bimodal firing-rate
distributions, being characterized by a negative excess
kurtosis, are preferred. 

We find robust performance and full homeostatic
adaption of the synaptic weights results as a by-product
of the synaptic flux minimization. This self-limiting behavior
allows for stable online learning for arbitrary durations.
The neuron acquires new information when the statistics of
input activities is changed at a certain point of the simulation,
showing however a distinct resilience to unlearn previously 
acquired knowledge. Learning is fast when starting with randomly
drawn synaptic weights and substantially slower when the
synaptic weights are already fully adapted. 

\tiny
 \keyFont{ \section{Keywords:} 
Hebbian learning \quad generating functionals \quad synaptic plasticity
\quad objective functions \quad Fisher information \quad homeostatic adaption}

\end{abstract}

\section{Introduction}
\label{Introduction}

Synaptic plasticity involves the modification of
the strength of individual synapses as a function of
pre- and postsynaptic neural activity. Hebbian
plasticity \citep{hebb2002organization} tends to
reinforce already strong synapses and may hence
lead, on a single neuron level, to runaway synaptic 
growth, which needs to be contained through 
homeostatic regulative processes \citep{turrigiano2000hebb},
such as synaptic scaling \citep{abbott2000synaptic}.
Modeling of these dual effects has been typically a
two-step approach, carried out by extending Hebbian-type 
learning rules by regulative scaling principles
\citep{bienenstock1982theory,oja1992principal,goodhill1994role,elliott2003analysis}.

An interesting question regards the fundamental computational
task a single neuron should be able to perform. There
is a general understanding that synaptic scaling induces
synaptic competition and that this synaptic competition 
generically results in a generalized principal component 
analysis (PCA) \citep{oja1992principal,miller1994role},
in the sense that a neuron will tend to align its vector 
of synaptic weights, within the space of input activities, 
with the direction having the highest variance. A meaningful
behavior, since information possibly transmitted by input 
directions with low variances is more susceptible to be 
obfuscated by internal or environmental noise.

A single neuron may however have additional computational
capabilities, in addition to its basic job as a principal 
component analyzer. The neuron may try to discover
`interesting directions', in the spirit of 
projection pursuit \citep{huber1985projection},
whenever the covariance matrix of the afferent inputs 
is close to unity. Deviations from Gaussian statistics
may encode in this case vitally important information, 
a well known feature of natural image statistics 
\citep{simoncelli2001natural,sinz2013temporal}.
One measure for non-Gaussianess is given by the 
kurtosis \citep{decarlo1997meaning}
and a single neuron may possibly tend to
align its synaptic weight vector with
directions in the space of input activities
characterized by heavy tails \citep{triesch2007synergies},
viz having a large positive excess kurtosis.
Here we study self-limiting Hebbian plasticity
rules which allow the neuron to discover maximally bimodal
directions in the space of input activities, viz directions
having a large negative excess kurtosis. 

Binary classification in terms of a linear discrimination 
of objects in the input data stream is a basic task for 
neural circuits and has been postulated to be a central 
component of unsupervised
object recognition within the framework of
slow feature analysis \citep{wiskott2002slow,dicarlo2012does}.
It is of course straightforward to train, using supervised
learning rules, a neuron to linearly separate the data 
received into two categories. Here we propose that a single
neuron may perform this task unsupervised, whenever it has
a preference for directions in the space of input activities 
characterized by negative excess kurtosis. Neural signals in 
the brain containing high frequency bursts have been linked to 
precise information transmission \citep{lisman1997bursts}. Neurons 
switching between relatively quiet and bursting states tend to 
have bimodal firing rate distributions and negative excess kurtosis.
The autonomous tendency to perform a binary classification, on
a single neuron level, may hence be of importance for higher
cortical areas, as neurons would tend to focus their intra-cortical
receptive fields towards intermittent bursting neural 
populations. A subclasss of bursting pyramidal neurons have been 
found in layer 5 of somatosensory and visual cortical areas 
\citep{chagnac1990bursts}. Neurons receiving input from these bursting 
cortical neurons would therefore be natural candidates to 
test this hypothesis, for which there is, to date, no direct 
experimental evidence. 

In order to develop synaptic plasticity rules, one may 
pursue one of two routes: either to reproduce certain 
aspects of experimental observations by directly formulating 
suitable plasticity rules, or to formulate, alternatively, 
an objective function from which adaption rules are 
then deduced \citep{intrator1992objective,bell1995information}.
Objective functions, also denoted generating functionals in the
context of dynamical system theory 
\citep{linkerhand2013generating,gros2014generating},
generically facilitate higher-level investigations, and have been used, e.g., 
for such as an overall stability analysis of Hebbian-type learning 
in autonomously active neural networks \citep{dong1992dynamic}.

The Fisher information measures the sensitivity of a system 
with respect to a given parameter. It can be related,
in the context of population coding \citep{brunel1998mutual},
to the transfer information between stimulus and neural
activity, and to order-parameter changes within the theory of
thermodynamic phase transitions \citep{prokopenko2011relating}.
Minimization of the Fisher information can be used as a generative
principle for quantum mechanics in general \citep{reginatto1998derivation}
and for the Euler equation in density functional theory 
\citep{nagy2003fisher}. Here we propose an objective function 
for synaptic learning rules based on the Fisher information 
with respect to a differential operator we denote the synaptic 
flux. 

The aim of adapting synaptic weights is to encode a maximal 
amount of information present in the statistics of the
afferent inputs. The statistics of the output neural activity
becomes stationary when this task is completed and the
sensitivity of the activity of the post-synaptic neuron
with regard to changes in the synaptic weights is then 
 minimal. Minimizing the Fisher information with respect
to the synaptic flux is hence a natural way to
generate synaptic plasticity rules. Morover, as we show in 
section \ref{Results}, the synaptic plasticity rules 
obtained by minimizing the Fisher information for the 
synaptic flux have a set of attractive features; 
incorporating standard Hebbian updating and being, at 
the same time, self-limiting. 

Minimizing an information theoretical objective function,
like the Fisher information, is an instance of polyhomeostatic
optimization \citep{markovic2010}, namely the optimization of
an entire function. Other examples of widely used 
information theoretical measures are the transfer entropy
\citep{vicente2011transfer} and the Kullback-Leibler divergence,
which one may use for adapting, on a slow time scale, 
intrinsic neural parameters like the bias, also called 
offset \citep{triesch2007synergies,markovic2012intrinsic}.
Minimizing the Kullback-Leibler divergence then corresponds to 
maximizing the information content, in terms of Shannon's information 
entropy, of the neural firing rate statistics. We use
intrinsic adaption for self-regulating the bias, obtaining,
as a side effect, an effective sliding threshold for the 
synaptic learning rule, in spirit of the BCM rule \citep{bienenstock1982theory}.

\section{Materials and Methods}
\label{Theory}

In the present work we consider rate encoding neurons for which  
the output firing rate $y$ is obtained as a sigmoidal function of the 
membrane potential $x$ via:
\begin{equation}
y = \sigma(x-b), \qquad
\sigma(z) = \frac{1}{1+\mathrm{e}^{-z}}, \qquad
x=\sum_{j=1}^{N_w} w_j (y_j-\bar y_j)~,
\label{eq_neuron_model}
\end{equation}
where $N_w$ is the number of input synapses, and $w_j$ and $y_j$ 
represent the synaptic weights and firing rates of the afferent 
neurons respectively. The sigmoidal $\sigma(z)$ has a fixed gain 
(slope) and the neuron has a single intrinsic parameter, the 
bias $b$. The $\bar y_j$ represent the trailing averages of $y_j$, 
\begin{equation}
\frac{d}{dt}\bar y_j = \frac{y_j-\bar{y}_j}{T_y} ~,
\label{eq_dot_y_bar}
\end{equation}
with $T_y$ setting the time scale for the averaging.
Synaptic weights may take, for rate encoding neurons, both 
positive and negative values and we assume here that
afferent neurons firing at the mean firing rate
$y_j\simeq \bar y_j$ do not influence the activity
of the postsynaptic neuron. This is a standard assumption
for synaptic plasticity which is incorporated in most
studies by appropriately shifting the mean of the 
input distribution.

In what follows we will derive synaptic plasticity rules for 
the $w_j$ and intrinsic plasticity rules that will optimize 
the average magnitude of $x$ and set in this way, implicitly, 
the gain of the transfer function. We have not included an 
explicit gain acting on $x$ since any multiplicative constant 
can be absorbed into the $w_j$ and, conversely, the average 
value of the $w_j$ can be thought of as the gain of the 
transfer function with rescaled $w_j$.

The firing rate $y$ of neurons has an upper limit,
an experimental observation which is captured by
restricting the neural output of rate encoding
neurons to the range $y\in[0,1]$. Here we consider with
\begin{equation}
F_{ob}(x,y) \ =\ E\left[\big(2+x\left(1-2y\right)\big)^2\right]
\label{eq_objectiveFunction}
\end{equation}
an objective function for synaptic plasticity rules
which treats the upper and the lower activity bounds 
on an equal footing. $E[\cdot]$ denotes the expectation 
value.

\begin{figure}[!t]
\begin{center}
\includegraphics[width=0.6\textwidth]{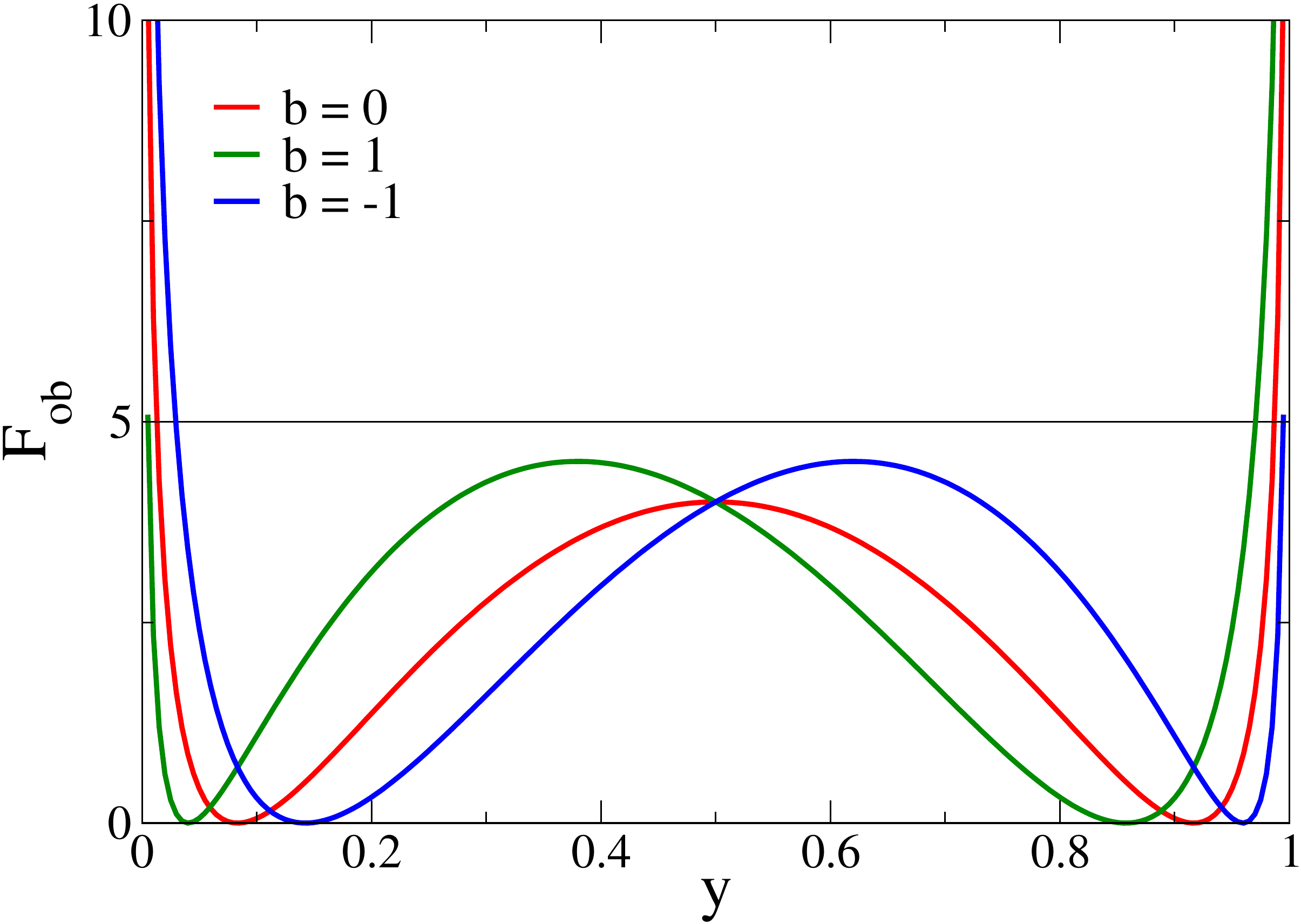}
\end{center}
\caption{The objective function $F_{ob}$, expression 
(\ref{eq_objectiveFunction}), as a function of the 
output firing rate $y$ for different values of the bias 
$b$. $F_{ob}$ always has two minima and diverges for 
extremal firing rates $y\to 0/1$, a feature responsible for inducing
output firing rates.
}
\label{F_ob}
\end{figure}

The functional $F_{ob}$ is positive definite and 
can be expressed as in (\ref{eq_objectiveFunction}),
or purely as a function of either $x$ or $y=\sigma(x-b)$. 
In Fig. \ref{F_ob}, $F_{ob}$ is plotted as a function of 
$y$ for different values of the bias $b$. The functional 
always presents two minima and diverges for extremal firing 
rates $0/1$. In particular, for firing rates $y\to0/1$, 
$F_{ob}$ is minimized by membrane potentials $x\to(-2)/2$, respectively.


Minimizing (\ref{eq_objectiveFunction}) as an objective 
function for deriving synaptic plasticity rules 
will therefore lead to bounded membrane potentials 
and hence necessarily to bounded learning rules, 
devoid of runaway synaptic growth. The cost function 
(\ref{eq_objectiveFunction}) generically has 
two distinct minima, a feature setting it apart from
other objective functions for synaptic plasticity rules
\citep{intrator1992objective}. Moreover, the objective 
function (\ref{eq_objectiveFunction}) can also be motivated 
by considering the Fisher information of the postsynaptic 
firing rate with respect to the synaptic flux, as shown in 
section \ref{sec_Fisher}. 

In section \ref{sec_flux_minimization}, via stochastic 
gradient descent, the following plasticity rules for 
the afferent synaptic weights $w_j$ will be derived:
\begin{equation}
\dot w_j \ =\ \epsilon_w G(x)H(x)(y_j-\bar{y}_j),
\label{eq_w_dot}
\end{equation}
with
\begin{equation}
G(x) = 2+x(1-2y), 
\qquad \quad H(x) = (2y-1)+2x(1-y)y. 
\label{eq_G_H}
\end{equation}
Here $\epsilon_w$ controls the rate of the synaptic 
plasticity. The bias $b$ entering the sigmoidal 
may be either taken to be constant or adapted via  
\begin{equation}
\dot b \ =\  (-\epsilon_b) \big (1-2y + y(1-y) \lambda\big),
\label{eq_b_dot}
\end{equation}
in order to obtain a certain average postsynaptic
firing rate, where $\lambda$ is a control parameter,
as detailed out in section \ref{sec_flux_minimization}.
Eq.~(\ref{eq_b_dot}) leads to the optimization of
the statistical information content of the neural activity, 
in terms of Shannon's information entropy; a process 
also denoted intrinsic adaption \citep{triesch2007synergies}
or polyhomeostatic optimization \citep{markovic2010}.

Both adaption rules, (\ref{eq_w_dot}) for the 
synaptic plasticity and (\ref{eq_b_dot}) for regulating
the average postsynaptic firing rate, interfere only
weakly. For instance one could take the bias $b$ 
as a control parameter, by setting $\epsilon_b\to0$, and
measure the resulting mean firing rate a posteriori. The 
features of the synaptic adaption process remain unaffected
and therefore alternative formulations for the intrinsic adaption 
of the bias could also be considered.

The synaptic plasticity rule (\ref{eq_w_dot})
involves the Hebbian factor $H(x)$, and a multiplicative
synaptic weight rescaling factor $G(x)$. Although here 
$G$ and $H$ are presented as a function of $x$ and $y$, 
these can also be expressed entirely in terms of $y$, 
consistently with the Hebbian interpretation. It is 
illustrative to consider the cases of small/large
postsynaptic neural activity. In the limit $y\to0/1$,
which is never reached, the updating rules
(\ref{eq_w_dot}) would read
\begin{equation}
\dot w_j \ \propto \
\left\{
\begin{array}{rcl}
 (2+x)\,(-1)\,(y_j-\bar y_j) && (y\to0)\\[0.5ex]
 (2-x)\,(+1)\,(y_j-\bar y_j) && (y\to1)
\end{array}
\right. .
\label{eq_flux-adaption-limits}
\end{equation}
For the case that $|x|<2$ we hence have that
the synaptic strength decreases/increases for
an active presynaptic neuron with $y_j>\bar y_j$, 
whenever the postsynaptic neuron is inactive/active,
an instance of Hebbian learning. The multiplicative 
constraint $(2\pm x)$ in (\ref{eq_flux-adaption-limits})
results in a self-limitation of synaptic growth. Synaptic
potentiation is turned into synaptic depression whenever 
the drive $x$ becomes too large in magnitude. Runaway 
synaptic growth is hence not possible and the firing rate
will settle close to the minima of $F_{ob}$, compare
Fig.~\ref{F_ob}.

\subsection{Motivation in terms of Fisher Information} 
\label{sec_Fisher}

The synaptic plasticity rules (\ref{eq_w_dot}) can be
derived either directly from the objective function
(\ref{eq_objectiveFunction}), as explained in 
section \ref{sec_flux_minimization} or motivated 
from an higher-order principle, the optimization 
of the synaptic flux, as we will show in the
following. Synaptic weight competition could be 
formulated, as a matter of principles, through 
an ad-hoc constraint like
\begin{equation}
\sum_j \left(w_j\right)^2  \to \mathrm{const.}, 
\qquad
\mathbf{w}=(w_1,w_2,\dots)~,
\label{eq_w_hypersphere}
\end{equation}
which defines a hypersphere in the phase of afferent
synaptic weights $\{w_j\}$, together with some appropriate
Hebbian-type adaption rules. We will not make use
of (\ref{eq_w_hypersphere}) explicitly, but our
adaption rules implicitly lead to finite length 
for the synaptic weight vector $\mathbf{w}$. 

Synaptic plasticity will modify, quite generically, the
statistical properties of the distribution $p(y)$
of the firing rate $y$ of the postsynaptic neuron.
It is hence appropriate to consider the sensitivity 
of the firing-rate distribution $p(y)$ with respect
to changes in the $w_j$. For this purpose one may
make use of the Fisher information
\begin{equation}
F_\theta  \ =\  \int  p(y)
\left(\frac{\partial}{\partial \theta}\ln\big(p(y)\big)\right)^2 
dy~,
\label{eq_fisher_info}
\end{equation}
which encodes the sensitivity of a given probability distribution
function $p(y)$ with respect to a certain parameter $\theta$.
Here we are interested in the sensitivity with respect to
changes in the synaptic weights $\{w_j\}$ and define with
\begin{equation}
F_w \  =\  \int  p(y)
\left(\sum_j w_j\frac{\partial}{\partial w_j}\ln\big(p(y)\big)\right)^2 
dy~,
\label{eq_F_flux}
\end{equation}
the Fisher information with respect to the synaptic flux.
Expression (\ref{eq_F_flux}) corresponds to the 
Fisher information (\ref{eq_fisher_info}) when
considering 
\begin{equation}
\frac{\partial}{\partial\theta} \ \to \
\sum_j w_j\frac{\partial}{\partial w_j}
 = \mathbf{w}\cdot\nabla_w,
\label{eq_flux_operator}
\end{equation}
as differential operator. The factors $w_j$ in front of the 
$\partial/\partial w_j$ result in a dimensionless expression,
the generating functional (\ref{eq_F_flux}) is then invariant 
with respect to an overall rescaling of the synaptic weights 
and the operator (\ref{eq_flux_operator}) a scalar. 
Alternatively we observe that 
$w_j\partial/\partial w_j =\partial/\partial\log(w_j/w_0)$,
where $w_0$ is an arbitrary reference synaptic weight,
corresponding to the gradient in the space of logarithmically 
discounted synaptic weights.

\begin{figure}[!t]
\begin{center}
\includegraphics[width=2.3in]{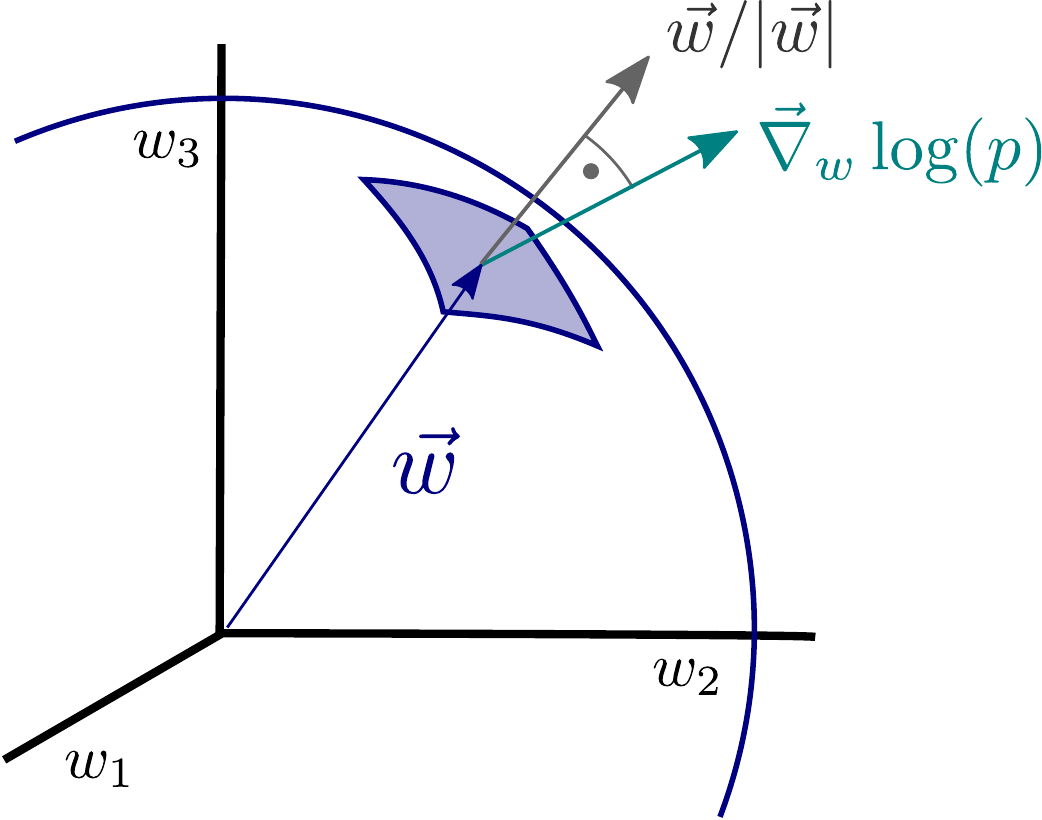}
\hspace{4ex}
\includegraphics[width=2.3in]{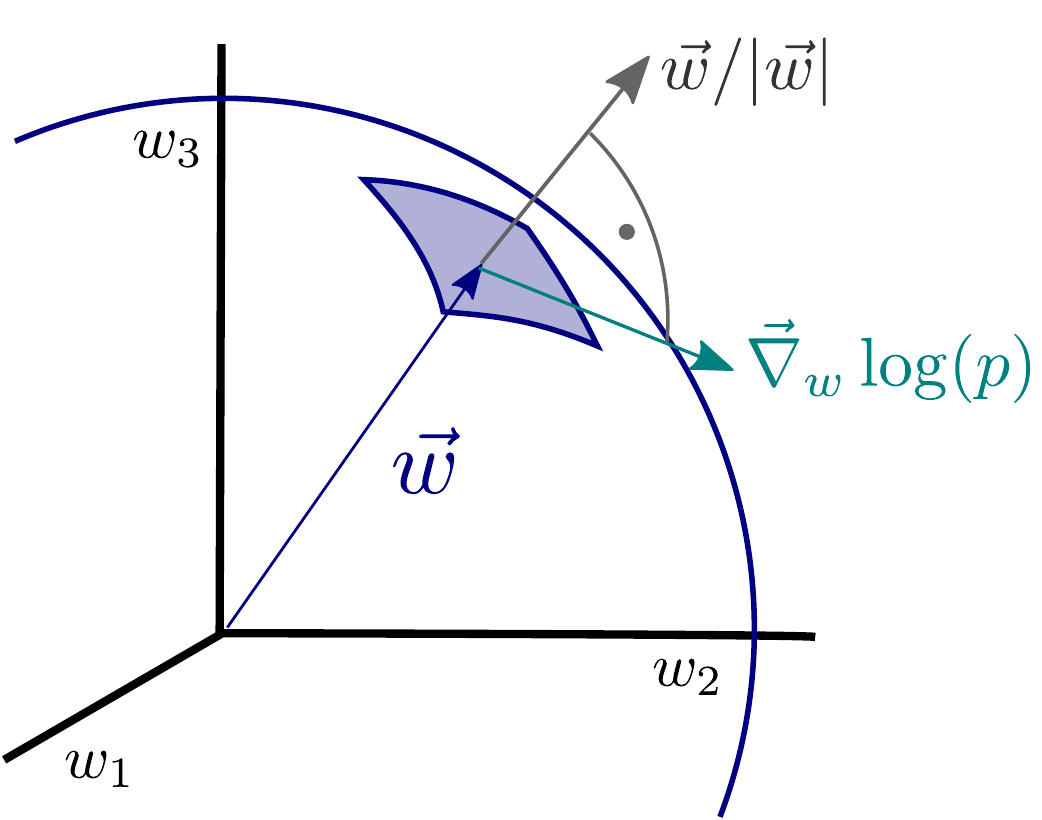}
\end{center}
\caption{
Illustration of the principle of minimal synaptic flux.
The synaptic flux, compare expression (\ref{eq_flux_operator}),
is the scalar product between the gradient 
$\nabla_{\mathbf{w}}\log(p)$ and the normal vector of the
synaptic sphere, $\vec w/|\vec w|$ (left). Here we disregard the
normalization. The sensitivity $\nabla_{\mathbf{w}}\log(p)$ of 
the neural firing-rate distribution $p=p(y)$, with respect 
to the synaptic weights $\mathbf{w}=(w_1,w_2,w_3,\dots)$, 
vanishes when the local synaptic flux is minimal (right), 
viz when $\mathbf{w}\cdot\nabla_{\mathbf{w}}\log(p)\to0$. 
At this point the magnitude of the synaptic weight 
vector $\vec w$ will not grow anymore.
}
\label{fig_flux}
\end{figure}

The operator (\ref{eq_flux_operator}), which we denote
{\em synaptic flux operator}, is, in addition, invariant 
under rotations within the space of synaptic weights and
the performance of the resulting synaptic plasticity rules 
will hence be also invariant with respect to the orientation
of the distributions $p(y_j)$ of the input activities $\{y_j\}$.
Physically, the operator (\ref{eq_flux_operator}) corresponds, 
apart from a normalization factor, to the local flux 
through the synaptic hypersphere, as defined by
Eq.~(\ref{eq_w_hypersphere}), since the synaptic vector
$\mathbf{w}$ is parallel to the normal vector through the 
synaptic hypersphere, as illustrated in Fig.~\ref{fig_flux}.


The Fisher information (\ref{eq_F_flux}) can be considered as
a generating functional to be minimized with respect to the 
synaptic weights $\{w_j\}$. The time-averaged properties of 
the neural activities, as measured by $p(y)$, will then not 
change any more at optimality; the sensitivity of the neural 
firing-rate distribution, with respect to the synaptic weights, 
vanishing for small $F_w$. At this point the neuron has 
finished encoding the information present in the input data 
stream through appropriate changes of the individual synaptic weights. 

It is interesting to consider what would happen if one would 
maximize the Fisher information instead of minimizing it. Then 
the neural firing activity would become very sensitive to small 
changes in the synaptic weights $\{w_j\}$ and information processing 
unstable, being highly susceptible to noise, viz to small statistical 
fluctuation of the synaptic weights. On a related note, the inverse 
Fisher information constitutes, via the Cramer-Rao theory 
\citep{paradiso1988theory,seung1993simple,gutnisky2008adaptive}, 
a lower bound for the variance when estimating an external parameter 
$\theta$. In this context, the external parameter $\theta$ can be 
estimated more reliably when the Fisher information is larger, viz when 
the distribution considered is highly sensible to the parameter of 
interest. This is a different setup. Here we are not interested in 
estimating the value of the synaptic weights, but in deducing adaption 
rules for the $\{w_j\}$.

\subsection{Synaptic flux minimization}
\label{sec_flux_minimization}

We are interested in synaptic plasticity rules which are 
instantaneous in time, depending only on the actual pre- and 
postsynaptic firing rates $y_j$ and $y$. Hence, the actual minimization 
of the synaptic flux functional (\ref{eq_F_flux}) needs to be 
valid for arbitrary distributions $p(y_j)$ of the presynaptic 
firing activities $\{y_j\}$. The synaptic flux $F_w$, which is 
in the first place a functional of the postsynaptic activity $p(y)$, 
needs therefore to be reformulated in terms of the distributions 
$p(y_j)$. A faithful representation of the postsynaptic firing-rate
distribution entering $F_w$ would involve a convolution over all 
presynaptic $p(y_j)$ and would hence lead to intricate cross-synaptic 
correlations \citep{bell1995information}. Our aim here, however, is to 
develop synaptic plasticity rules for individual synapses, functionally 
dependent only on the local presynaptic activity and on the overall
postsynaptic firing level. We hence consider for the minimization of 
the synaptic flux all $j\in\{1,\dots,N_w\}$ synapses separately, viz we replace 
(\ref{eq_F_flux}) by
\begin{eqnarray}
F_w  &\to&  \int \left(\sum_j w_j\frac{\partial}{\partial w_j}
\ln\left(\frac{p(y_j)}{\partial y/\partial y_j}\right)\right)^2 
\prod_l p(y_l) dy_l \nonumber \\
&\equiv & \int  f_w(y) \prod_l p(y_l)dy_l~,
\label{eq_F_flux_y_j}
\end{eqnarray}
where we have defined the kernel $f_w(y)$. We denote the approximation 
(\ref{eq_F_flux_y_j}) the {\em local synapse approximation},
since it involves the substitution of $p(y)dy$ by $ \prod_l p(y_l)dy_l$. 
Expression (\ref{eq_F_flux_y_j}) becomes exact for the case $N_w=1$. 
It corresponds to the case in which the distinct afferent synapses 
interact only via the overall value of the membrane potential $x$,
as typical for a mean-field approximation.
We then find, using the neural model (\ref{eq_neuron_model}),
\begin{equation}
\frac{\partial y}{\partial y_j} \ =\ y(1-y)w_j
\label{eq_partial_y_partial_y_j}
\end{equation}
and hence
\begin{equation}
f_w(y) \ = \  
\left(\sum_j w_j\left(\frac{1}{w_{j}}+(y_j-\bar y_j)(1-2y) \right) \right)^2 
\ =\ \Big(N_w+x(1-2y) \Big)^2 
\label{eq_F_w_kernel}
\end{equation}
where $N_w$ is the number of afferent synapses.
The kernel $f_w$ is a function of $y$ only, and not of
the individual $y_j$, since $x=\sum_j w_j(y_j-\bar y_j)$.
More fundamentally, this dependency is a consequence of
choosing the flux operator (\ref{eq_flux_operator})
to be a dimensionless scalar. 

Taking $N_w\to2$ in (\ref{eq_F_w_kernel}) leads to the 
objective function (\ref{eq_objectiveFunction}) and 
results in $G(x)$ and $H(x)$ being proportional 
to each other's derivatives, with the roots and maxima 
respectively aligned. We however also performed simulations 
using the generic expression (\ref{eq_F_w_kernel}), with 
the results changing only weakly and quantitatively. 

The synaptic weights are updated so that $f_w(y)$ 
becomes minimal, $\dot w_j  \propto - \partial f_w(y)/\partial w_j$, 
obtaining the plasticity rule (\ref{eq_w_dot}). 
This procedure corresponds to a stochastic steepest 
descent of the objective function (\ref{eq_objectiveFunction}),
a procedure employed when one is interested in obtaining
update rules independent of the actual distributions
$p(y_j)$ of the afferent neural activities. 

For the derivation one 
makes use of $\partial y/\partial w_j = (y_j-\bar y_j)(1-y)y$.
The synaptic plasticity rule (\ref{eq_w_dot}) depends via
$y_j-\bar y_j$ on the activity $y_j$ of the presynaptic neuron 
relatively to its mean firing rate $\bar y_j$. This dependence
models experimental findings indicating that, in the context of spike 
timing dependent plasticity, low-frequency stimulation generically 
induces causal depression 
\citep{shouval2010spike,lisman2010questions,feldman2012spike};
one needs above-average firing rates for causal potentiation.

Note that synaptic competition is present implicitly in the updating rule
through the membrane potential $x$, entering both $G(x)$ and $H(x)$,
which integrates all individual contributions; the local synapse 
approximation (\ref{eq_F_flux_y_j}) only avoids explicit cross synaptic 
learning.

We denote the two factors on the right-hand side of (\ref{eq_w_dot}), 
$G(x)$ and $H(x)$, as self-limiting and Hebbian respectively;
with $H(x)$ being, by construction, the derivative of $G(x)$. 
The derivative of $H(x)$ is also proportional to $G(x)$
since we substituted $N_w\to2$ in the objective function
on the right-hand-side of Eq.~(\ref{eq_F_w_kernel}). With this
choice, the two factors $G(x)$ and $H(x)$ are hence conjugate
to each other.

The synaptic plasticity rule (\ref{eq_w_dot}) works robustly for 
a wide range of adaption rates $\epsilon_w$, including the case 
of online learning with constant updating rates. For all simulations 
presented here we have used $\epsilon_w=0.01$. We constrained the
activities of the presynaptic neurons $y_j$, for consistency,
to the interval $[0,1]$, which is the same interval of postsynaptic
firing rates. Generically we considered uni- and bi-modal Gaussian 
inputs centered around $\bar y_j=0.5$, with individual standard 
deviations $\sigma_j$. We considered in general $\sigma_j=0.25$ 
for the direction having the largest variance, the dominant direction, 
with the other directions having smaller standard deviations, 
typically by a factor of two.

\subsection{Emergent sliding threshold}

\begin{figure}[!t]
\begin{center}
\includegraphics[width=0.45\textwidth]{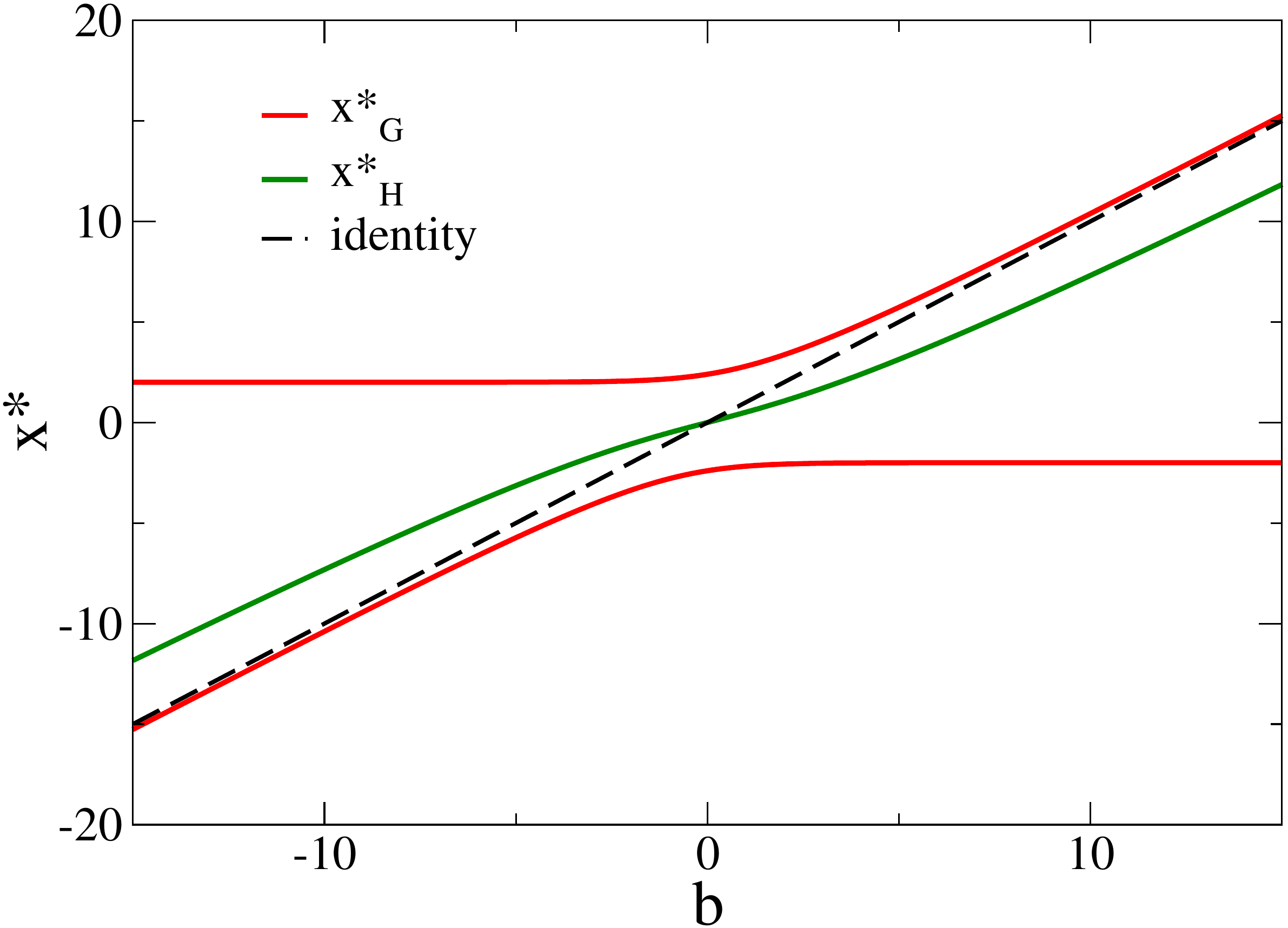}\hspace{1ex}
\includegraphics[width=0.45\textwidth]{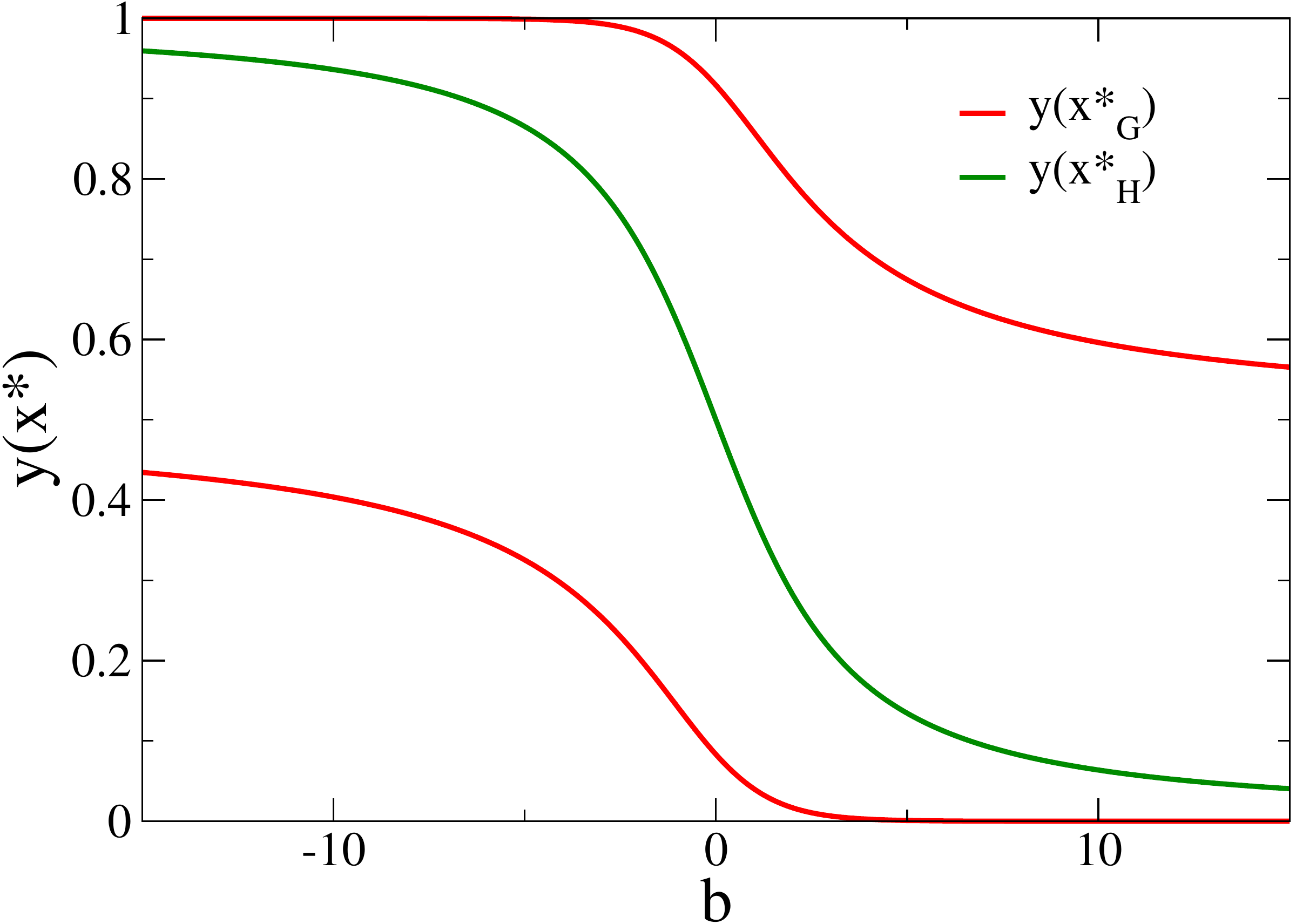}
\end{center}
\caption{
{The roots of the adaption factors.} 
Left: The roots $G(x^*_{0,1})=0$ and $H(x^*)=0$ respectively,
compare Eqs.~(\ref{eq_G_H}) and (\ref{eq_G_H}), as
a function of the bias $b$. Note that the roots do 
not cross, as the factors $G$ and $H$ are conjugate 
to each other.
Right: The respective values $y(x^*)$ of the neural 
activity. Note that $y(x_1^*)-y(x_0^*)\ge 1/2$, for 
all values of the bias. 
}
\label{fig_GH_roots}
\end{figure}

One may invert the sigmoidal $\sigma(x)$ via
$x=b-\log\big((1-y)/y\big)$ and express the adaption
factors solely in terms of the neural firing rate $y$.
For the Hebbian factor $H(x)$, see Eq.~(\ref{eq_G_H}),
one then finds
\begin{equation}
H(y)  = (2y-1)\,+\, 2y(1-y)\left[b-\log\big((1-y)/y\big)\right]~.
\label{eq_H_y}
\end{equation}
The bias $b$ hence regulates the crossing point from anti-Hebbian 
(for low neural activity $y<y^*_H$) to Hebbian learning 
(for large firing rates $y>y^*_H$), where $y^*_H$ is the root of 
$H(y)$. $y^*_H$ depends only on $b$ 
(as shown in Fig.~\ref{fig_GH_roots}), emerging then indirectly 
from the formulation of the objective function 
(\ref{eq_objectiveFunction}), and plays the role of a sliding
threshold. This sliding threshold is analogous to the one
present in the BCM theory \citep{bienenstock1982theory}, 
which regulates the crossover from anti-Hebbian to Hebbian
learning with increasing output activity and which is 
adapted in order to keep the output activity within a
given working regime.


The bias $b$ regulates, in addition to its role determining the 
effective sliding threshold for synaptic plasticity, the mean 
firing rate. In principle, one may consider an ad-hoc update rule 
like $\dot b\propto (\mu-\bar y)$ for the bias, where 
$\mu=\int y p_\lambda(y) dy$ is some given target firing 
rate and where $\bar y$ would be a sliding average of $y$. 
We will however use, alternatively, an information theoretical 
objective function for the intrinsic adaption of the bias. 
The Kullback-Leibler divergence 
\begin{equation}
D  = \int dy\, p(y)\log\left(\frac{p(y)}{p_\lambda(y)}\right),
\qquad p_\lambda(y) = \frac{\mathrm{e}^{\lambda y}}{N_\lambda}
\label{eq_D_KL}
\end{equation}
measures the distance between the actual firing-rate distribution
$p(y)$ and a given target distribution $p_\lambda(y)$. It will
be minimal if $p_\lambda(y)$ is approximated as well as possible.
An exponential target distribution, as selected here, maximizes
the information content of the neural activity in terms
of information entropy, given the constraint of a fixed mean $\mu$, 
both for a finite support $y\in[0,1]$, as considered here,
as well as for an unbounded support, $y>0$, with $N_\lambda$ 
being the appropriate normalization factor. For $\lambda\to0$ 
a uniform target distribution is recovered together with 
$\mu\to0.5$ and the resulting $p(y)$ becomes symmetric with
respect to $y=0.5$.

Following a derivation which is analogous to the one given 
above for the case of synaptic flux minimization, one 
finds Eq.~(\ref{eq_b_dot}) for the adaption rules 
\citep{triesch2007synergies,linkerhand2013self}.
For the adaption rate $\epsilon_b$ for the bias 
we used in our simulations generically $\epsilon_b=0.1$,
its actual value having only a marginal influence on 
the overall behavior of the adaption processes.

Minimizing the Kullback-Leibler divergence and
the Fisher information are instances of 
polyhomeostatic optimization
\citep{markovic2010,markovic2012intrinsic}, as one 
targets to optimize an entire probability distribution
function, here $p(y)$. An update rule like 
$\dot b\propto (\mu-\bar y)$ would, on the other side, 
correspond to a basic homeostatic control, aiming to 
regulate a single scalar quantity, such as the mean 
firing rate.

\subsection{Fixpoints of the limiting factor}

The self-limiting factor $G(x)$ has two roots $x^*_G$,
compare Fig.~\ref{fig_GH_roots}. For $b=0$ one finds 
$x^*_G\approx\pm 2.4$ corresponding to firing-rates 
$y^*_G=0.083$ and $y^*_G=0.917$ respectively, 
compare also Fig.~\ref{fig_GH_roots}. The roots
of $G(x)$ are identical with the two minima of the
objective function $F_{ob}$, compare (\ref{eq_objectiveFunction}).
The self-limiting nature of the synaptic adaption rules
(\ref{eq_w_dot}) is a consequence of the two roots of $G(x)$, 
as larger (in magnitude) membrane potentials will reverse the 
Hebbian adaption to an anti-Hebbian updating. The roots of
$G(x)$ induce, in addition, the tendency of performing
a binary classification. As an illustration consider the case 
of random sequences of discrete input patterns
\begin{equation}
\mathbf{y}^\eta,
\qquad\quad \eta =1,\dots,N_{patt}~, 
\label{eq_y_in_gamma}
\end{equation}
with the number of input patterns $N_{patt}$ being
smaller than the number of afferent neurons,
$N_{patt}\le N_w$. The inputs $(y_1,\dots,y_{N_w})=\mathbf{y}$ 
are selected randomly out of the set (\ref{eq_y_in_gamma})
of $N_{patt}$ patterns and presented consecutively for
small time intervals. The synaptic updating rules will then
lead, as we have tested through extended simulations, to a
synaptic vector $\mathbf{w}$ dividing the space of input
patterns into two groups, 
\begin{equation}
\begin{array}{rclrrcr}
\mathbf{w}\cdot\left(\mathbf{y}^\eta-\mathbf{\bar y}\right) &=& x^*_G(1) &
\qquad  \mathrm{for} & \gamma N_{patt} & \mathrm{states} & \mathbf{y}^\eta\\[0.5ex]
\mathbf{w}\cdot\left(\mathbf{y}^\eta-\mathbf{\bar y}\right) &=& x^*_G(2) &
\qquad  \mathrm{for} & (1-\gamma) N_{patt} &\mathrm{states} & \mathbf{y}^\eta
\end{array}~,
\label{eq_binary_classifier}
\end{equation}
which is a solvable set of $N_{patt}$ equations for $N_w$ 
variables $(w_1,w_2,\dots)$. Here we have denoted with
$x^*_G(1)$ and $x^*_G(2)$  the two distinct roots of $G(x)$,
and with
$\mathbf{\bar y}=\left(\sum_\eta\mathbf{y}^\eta\right)/N_{patt}$
the mean input activity.
This outcome of the long-term adaption corresponds to 
a binary classification of the $N_{patt}$ vectors. The
membrane potential $\mathbf{x}=\mathbf{w}\cdot\mathbf{y}$
just takes two values, for all inputs $\mathbf{y}$ drawn
from the the set of input patterns.  

There is one free parameter in (\ref{eq_binary_classifier}), 
namely the fraction $\gamma$ and $(1-\gamma)$ of patterns 
mapped to $x^*_G(1)$ and $x^*_G(2)$ respectively. This
fraction $\gamma$ is determined self-consistently by the system,
through the polyhomeostatic adaption (\ref{eq_b_dot}) of 
the bias $b$, with the system trying to approximate as close
as possible the target firing-rate distribution 
$\propto\mathrm{exp}(\lambda y)$, see Eq.~(\ref{eq_D_KL}).

\section{Results}
\label{Results}

In order to test the behavior of the neuron under 
rules (\ref{eq_w_dot},\ref{eq_b_dot}) when presented 
with different input distributions, a series of 
numerical simulations have been performed. In the 
following sections, the evolution of the system when 
faced with static input distributions is first studied. 
In particular, principal component extraction and linear 
discrimination tasks are evaluated. These results are 
then extended to a scenario of varying input distributions 
and a fading memory effect is then analyzed. 

\subsection{Principal component extraction}

\begin{figure}[!t]
\begin{center}
\begin{tabular}{cr}
\raisebox{1ex}{\bf (A)}\includegraphics[width=0.26\textwidth]{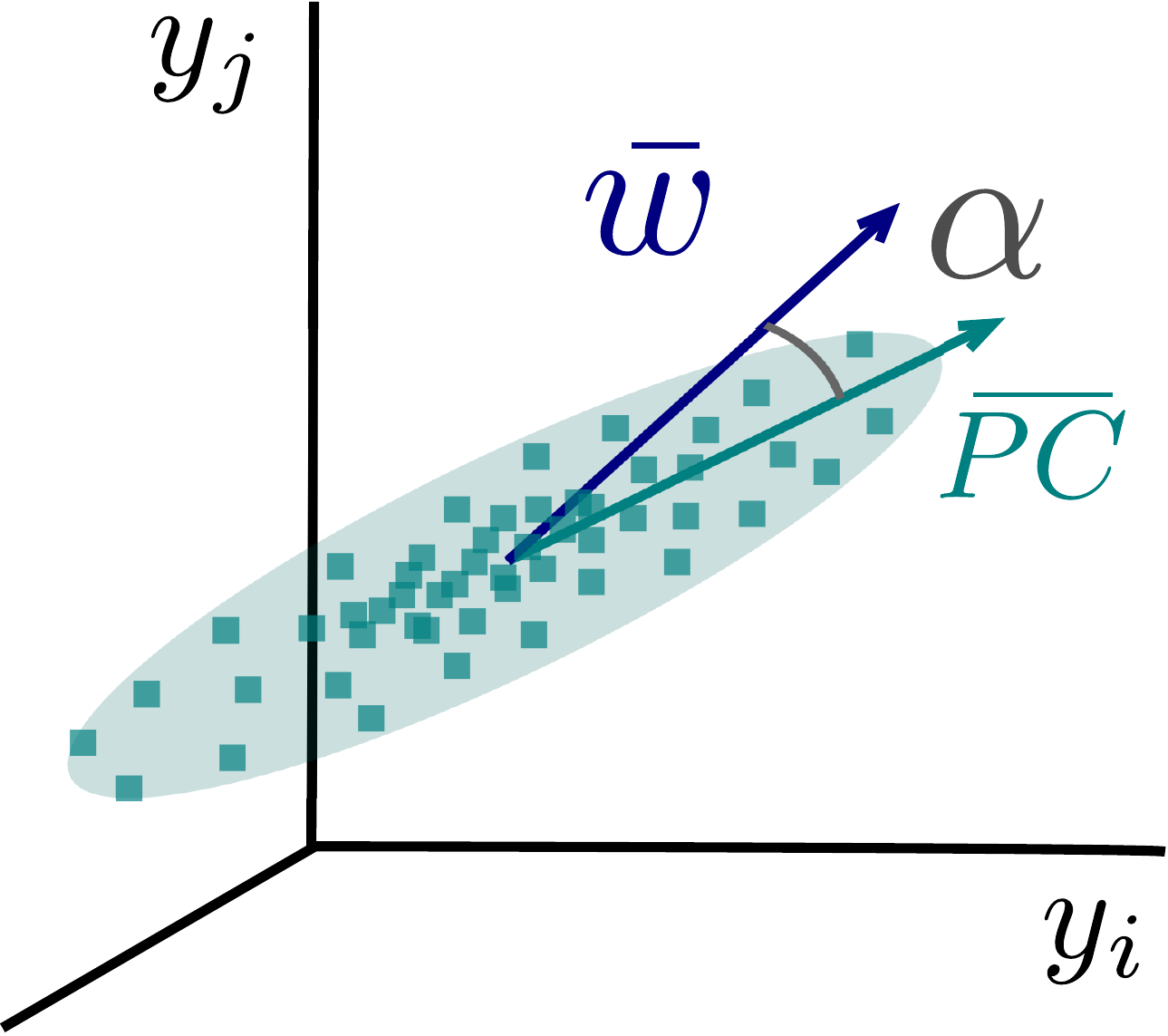}&
\raisebox{1ex}{\bf (B)}\includegraphics[width=0.37\textwidth]{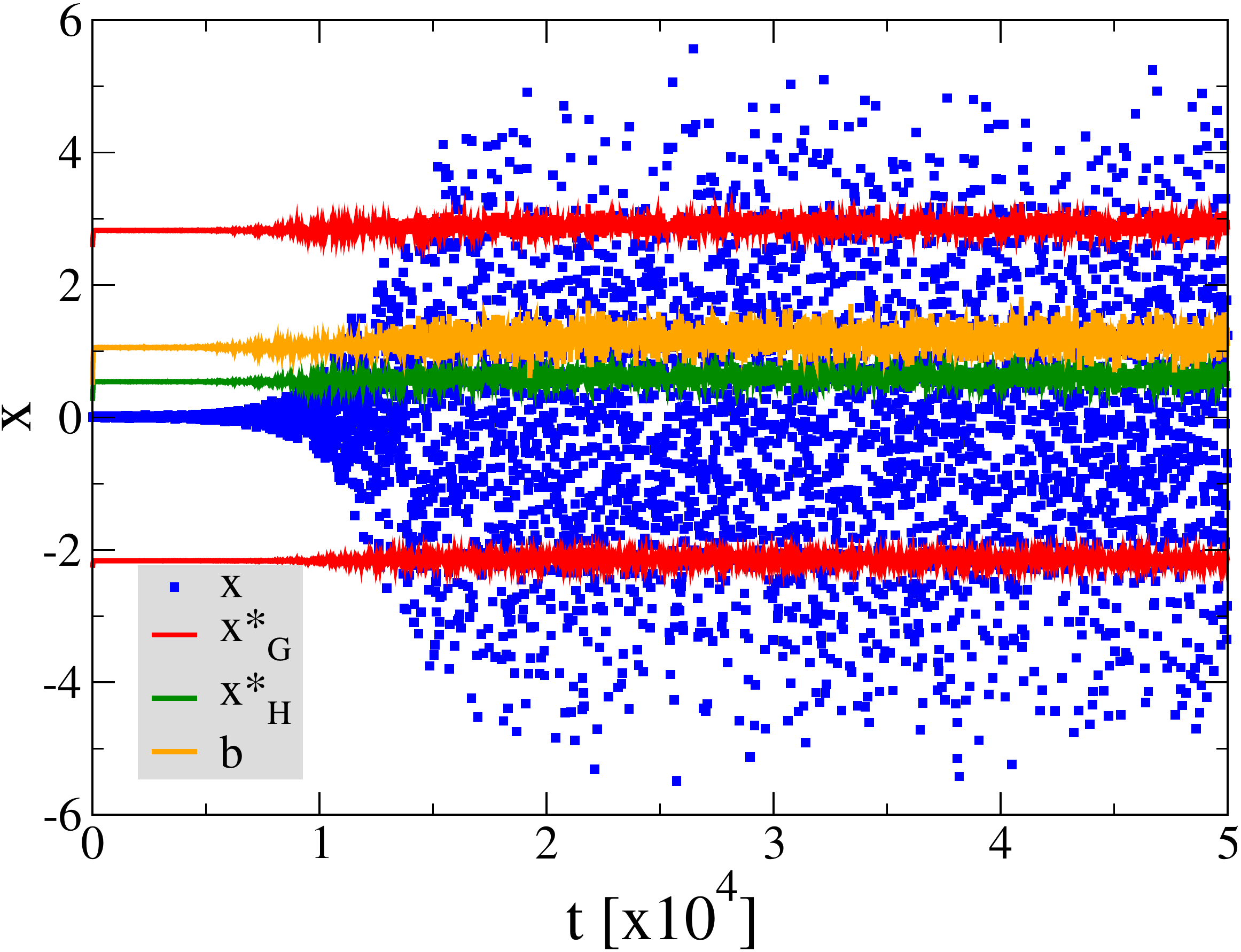}\\
\raisebox{1ex}{\bf (C)}\includegraphics[width=0.32\textwidth]{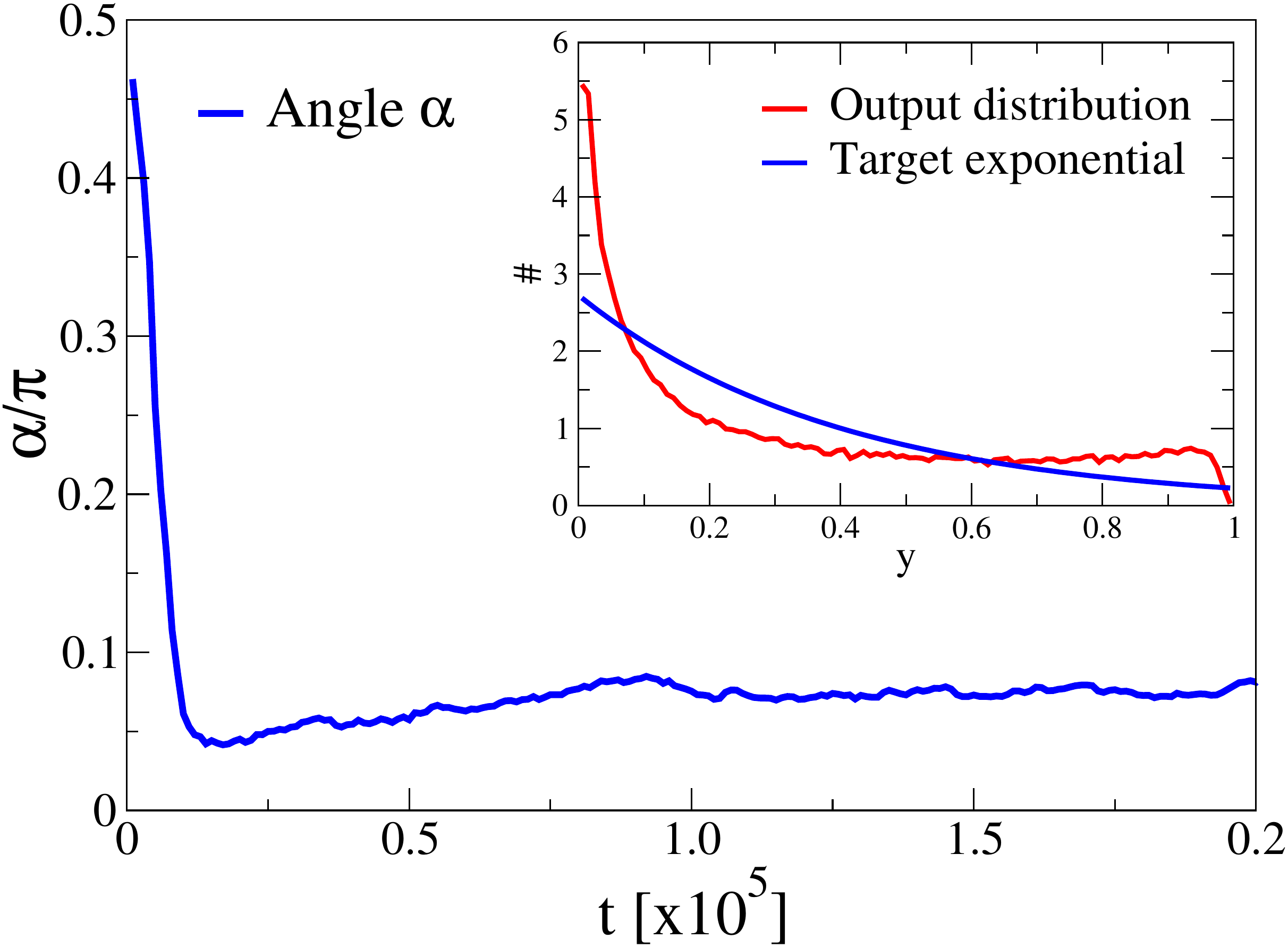}&
\raisebox{1ex}{\bf (D)}\includegraphics[width=0.37\textwidth]{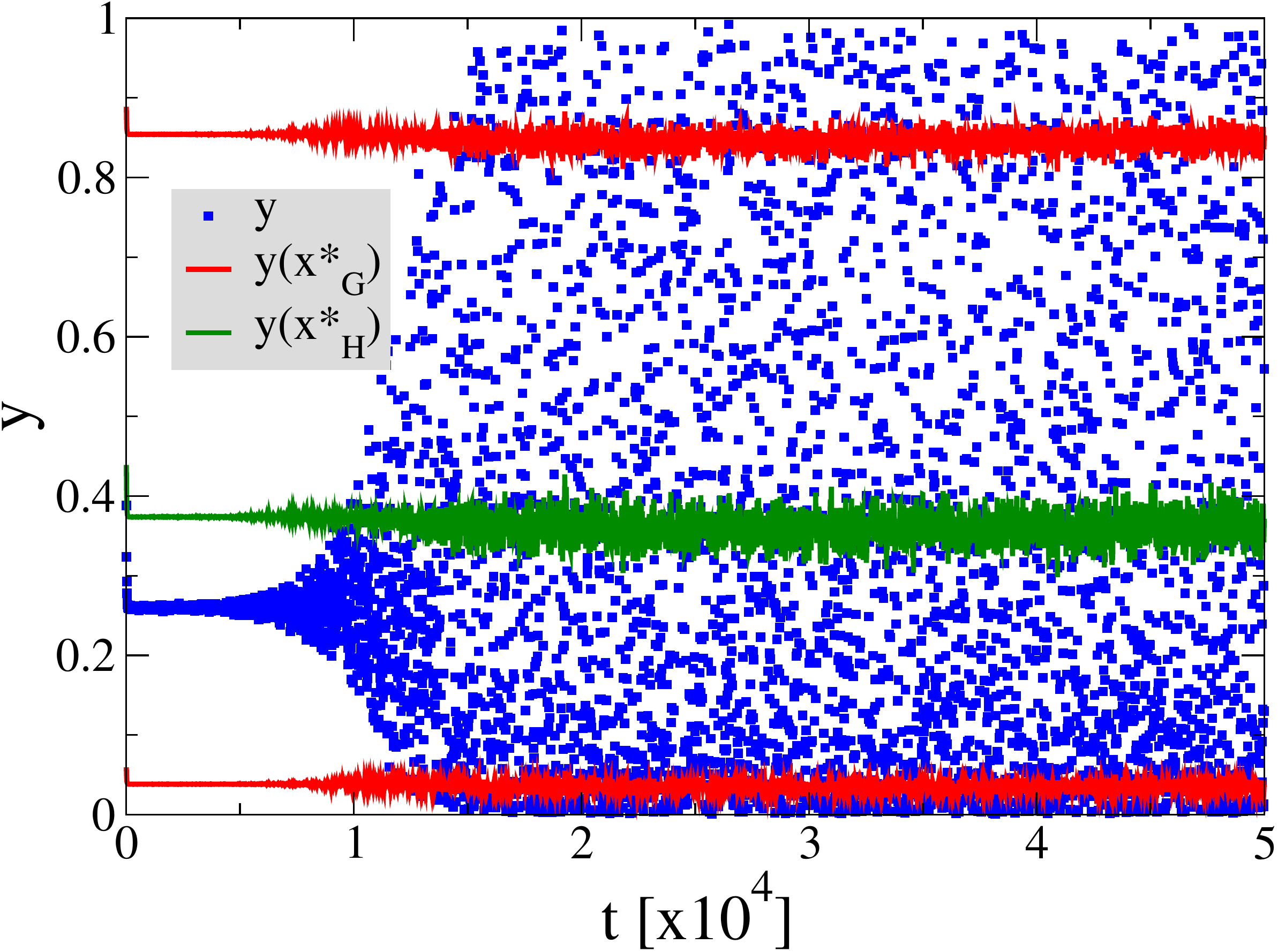}\\
\raisebox{1ex}{\bf (E)}\includegraphics[width=0.32\textwidth]{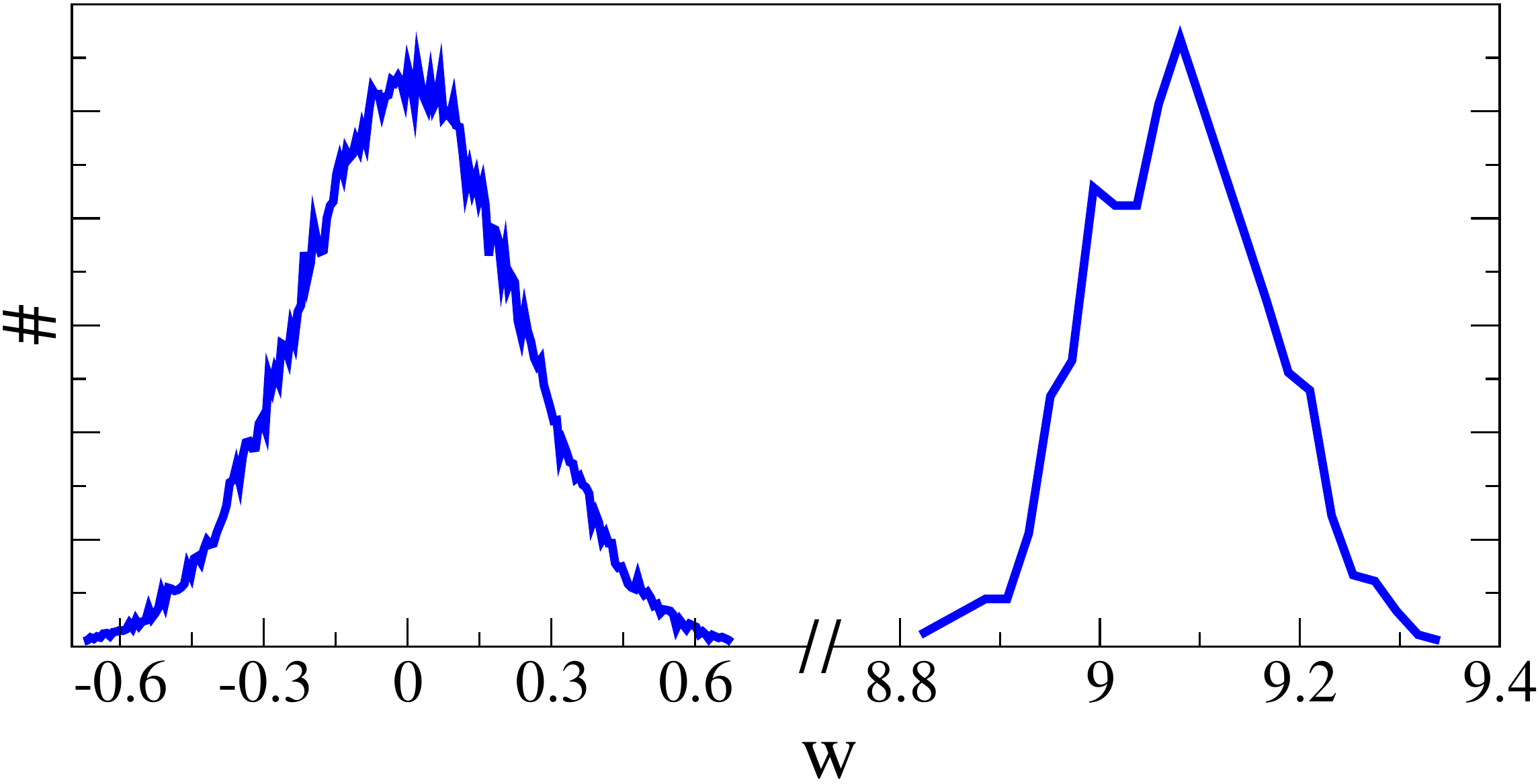}&
\raisebox{1ex}{\bf (F)}\includegraphics[width=0.37\textwidth]{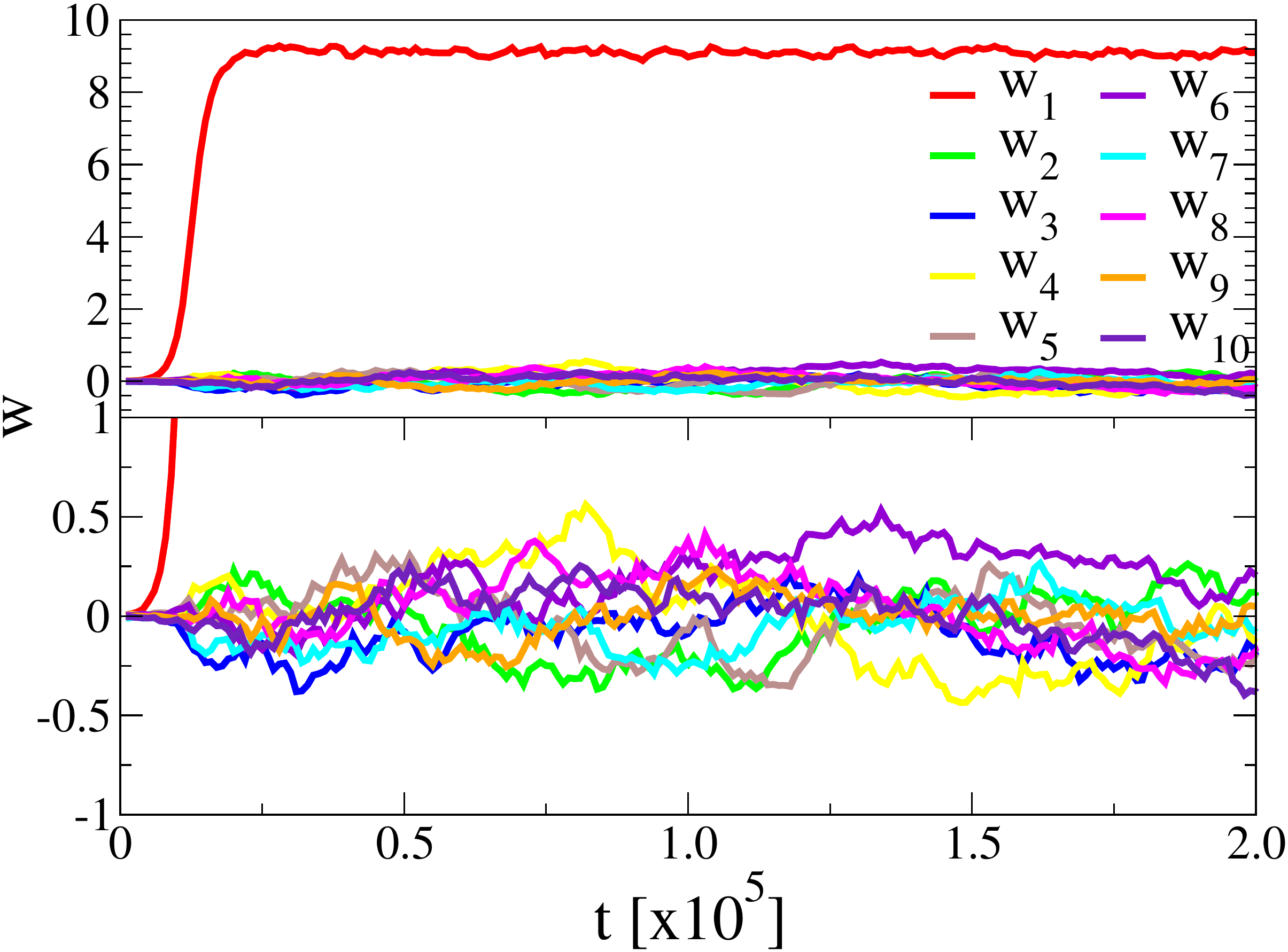}
\end{tabular}
\end{center}
\caption{
{Alignment to the principal component.} 
Simulation results for a neuron with $N_w=100$
input neurons with Gaussian  input distributions
with one direction (the principal component) having twice the 
standard deviation than the other $N_w-1$ directions.
\textbf{(A)} 
Illustration of the input distribution density $p(y_1,y_2,\dots)$, with
the angle $\alpha$ between the direction of the
principal component ($\overline{PC}$) and $\overline{w}$, the synaptic weight vector.
\textbf{(B)} 
Time series of the membrane potential $x$ (blue), the bias $b$ (yellow),
the roots $x^*_G$ of the limiting factor $G(x)$ (red) and the
root $x^*_H$ of the Hebbian factor $H(x)$ (green).
\textbf{(C)} 
The evolution of the angle $\alpha$ of the synaptic weight vector 
$\mathbf{w}$ with respect to the principal component and 
(inset)
the output distribution $p(y)$ (red) with respect to the target
exponential (blue).
\textbf{(D)} 
Time series of the output $y$ (blue) and of the roots
$y^*_G$ of the limiting factor $G(y)$ (red) and the
root $y^*_H$ of the Hebbian factor $H(y)$ (green).
\textbf{(E)} 
Distribution of synaptic weights $p(w)$ in the stationary
state for large times.
\textbf{(F)} 
Time evolution of the first ten synaptic weights $\{w_j\}$,
separately for the principal component (upper panel) and
for nine other orthogonal directions (lower panel).
}
\label{fig_PCA}
\end{figure}

As a first experiment we consider the case of $N_w$
input neurons with Gaussian activity distributions
$p(y_j)$. In this setup a single component, namely
$y_1$, has standard deviation $\sigma$ and all other $N_w-1$
directions have a smaller standard deviation of $\sigma/2$, as
illustrated in Fig.~\ref{fig_PCA}(A). We have selected, 
for convenience, $y_1$ as the direction of the principal 
component. The synaptic updating rule (\ref{eq_neuron_model})
is however fully rotational invariant in the space
of input activities and the results of the simulations
are independent of the actual direction of the 
principal component. We have verified this independence
by running simulations with dominant components selected 
randomly in the space of input activities.


In Fig.~\ref{fig_PCA} we present the result for
$N_w=100$ afferent neurons and $\lambda=-2.5$
for the target distribution $p_\lambda(y)$, compare 
Eq.~(\ref{eq_D_KL}) in section \ref{Theory}.
The initial synaptic weights $\{w_j\}$ have
been randomly drawn from $[-0.005:0.005]$ and 
are hence quite small, such that the learning rule 
is initially exclusively Hebbian,
viz the membrane potential $x$ is substantially
smaller than the roots $x^*_G$ of the limiting
factor $G(x)$ (compare Fig.~\ref{fig_PCA}(B) where
$x$/$x^*_G$ are given by the blue/red dots respectively).
Hebbian synaptic growth then eventually leads
to larger weights, with the weight along the
principal component (here $w_1$, red line in  
Fig.~\ref{fig_PCA}(F)) becoming very large.
At this stage, the membrane potential $x$ starts to cross
the roots $x^*_G$ of the limiting factor $G(x)$ and a stationary
state results, with the weight along the principal component
saturating and with the weights along the non-principal components
involved in bounded random drifts. This stationary state, with 
continuously ongoing online learning, remains stable for arbitrary 
simulation times.

The firing rate $y(t)$ covers the whole available interval 
$[0,1]$, in the stationary state, and a sliding threshold emerges 
self-consistently. This sliding threshold is given by the root 
$x^*_H$ of the Hebbian factor $H(x)$; learning is Hebbian/anti-Hebbian
for $y>y(x^*_H)$ and  $y<y(x^*_H)$ respectively. For our simulation the
sliding threshold is about $y(x^*_H)\simeq0.4$ 
(green dots in Fig.~\ref{fig_PCA}(D)) in the stationary state.

The angle $\alpha$ between the direction of the synaptic weight
vector $\mathbf{w}$ and the principal component of input activities
is initially large, close to the random value of $\pi/2$, dropping
close to zero with forthgoing synaptic adaption, as shown in
Fig.~\ref{fig_PCA}(C), a consequence of the growth of $w_1$. In
Fig.~\ref{fig_PCA}(E) we plot the distribution of the $w_j$, with
a separate scale for the principal component, here $w_1\approx 9.1$ (as 
averaged over 100 runs). The small components are Gaussian distributed 
around zero with a standard deviation of $\sigma_w^{(non)} \approx 0.23$, 
we have hence a large signal-to-noise ratio of 
$S_w = |w_1|/\sigma_w^{(non)}\approx 9.1/0.23\approx 40$.

We also present in the inset of Fig.~\ref{fig_PCA}(C) a 
comparison between the actual firing-rate distribution 
$p(y)$ in the stationary state and the exponential target 
distribution $\propto\exp(\lambda y)$, entering the
Kullback-Leibler divergence, see Eq.~(\ref{eq_D_KL}).

\begin{figure}[!t]
\begin{center}
\includegraphics[width=0.4\textwidth]{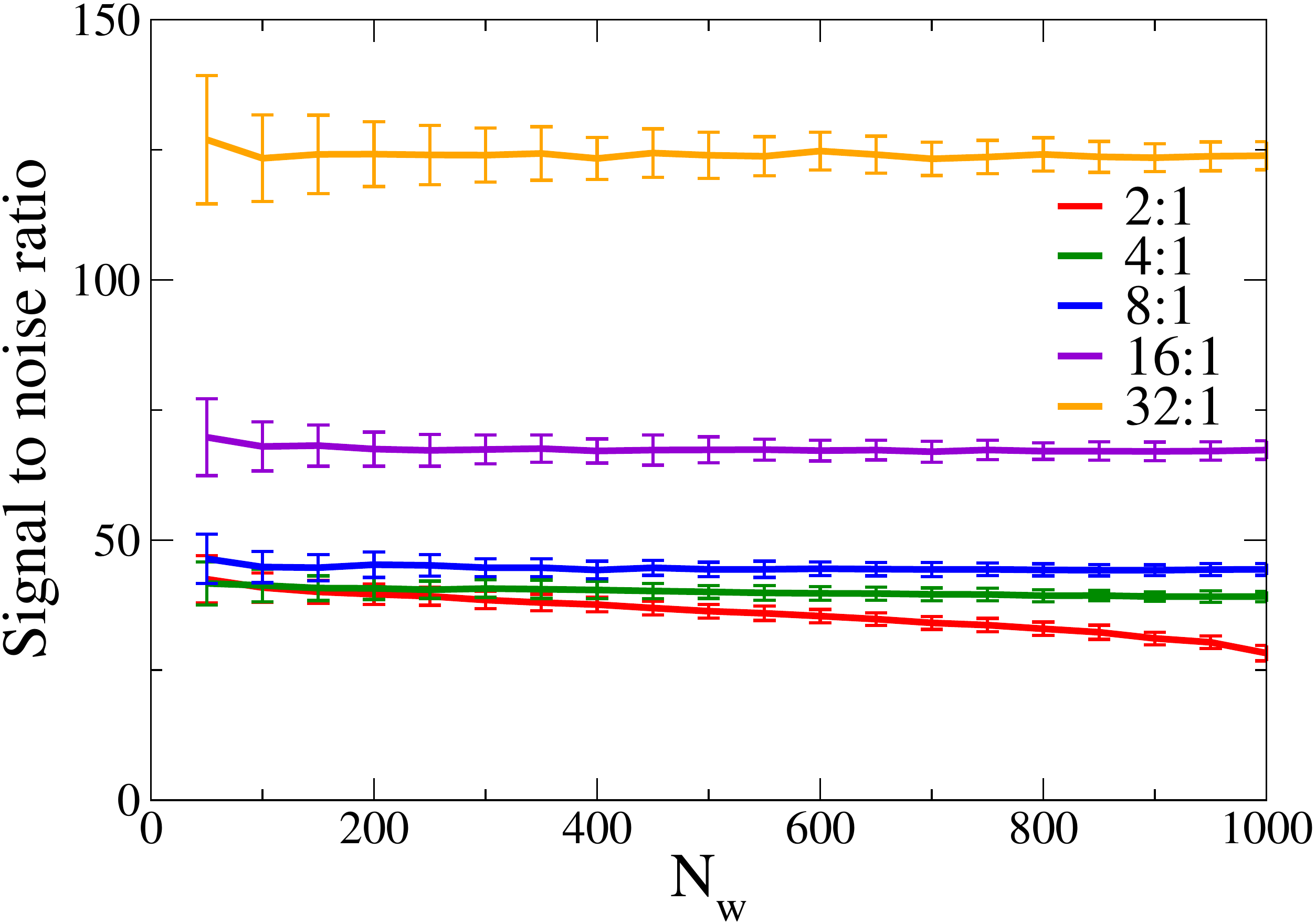}
\hspace{1ex}
\includegraphics[width=0.4\textwidth]{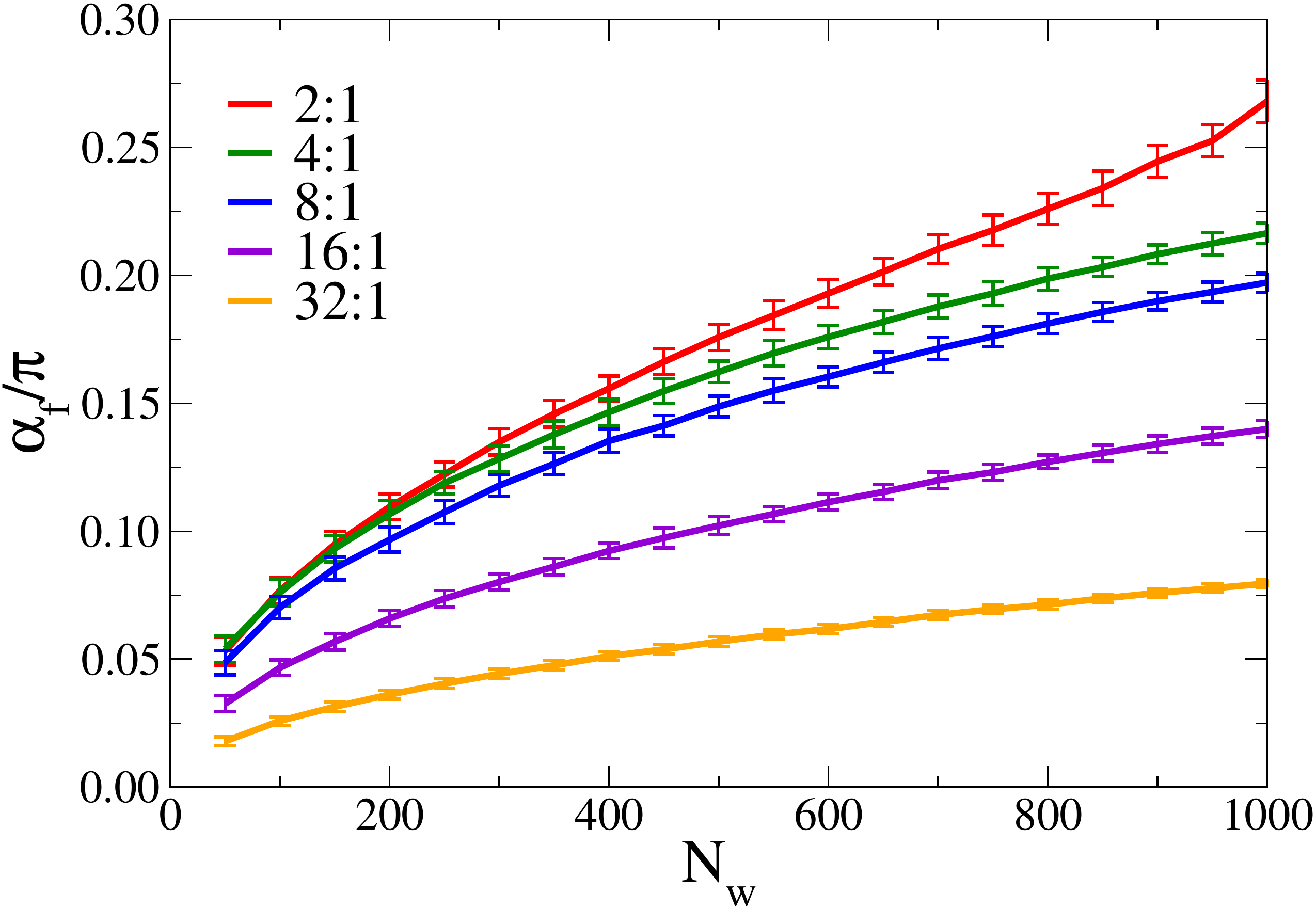}
\end{center}
\caption{
{Scaling of the adaption rules with the number of afferent neurons.} 
For constant simulating parameters the signal to noise ratio (left), 
defined as the ratio $|w_1|/\sigma_w^{(non)}$, where $w_1$ is 
the synaptic strength parallel to the principal component and 
$\sigma_w^{(non)}$ the standard deviation of the orthogonal synaptic 
directions, compare Eq.~(\ref{eq_alpha_N_w}), and the mean angle (right),
of the synaptic weight vector with respect to the principal component. 
Shown are results for a range, 2:1, 4:1, 8:1, 16:1 and 32:1,
of the incoming signal-to-noise ratios, defined as the
ratio of the standard deviations between the large and the small 
components of the distributions of input activities $p(y_j)$. 
The outgoing signal-to-noise ratio $|w_1|/\sigma_w^{(non)}$
remains essentially flat, as a function of $N_w$; the increase 
observed for the average angle $\alpha$ is predominantly a statistical 
effect, caused by the presence of an increasingly large number 
of orthogonal synaptic weights. The orthogonal weights are all individually 
small in magnitude, but their statistical influence sums up 
increasingly with raising $N_w$.
        }
\label{fig_N_scaling}
\end{figure}

\subsection{Signal-to-noise scaling}

For synaptic adaption rules to be biologically significant
they should show stable performance even for large numbers $N_w$
of afferent neurons, without the need for fine-tuning of the
parameters. This is the case for our plasticity rules.

In Fig.~\ref{fig_N_scaling} we present the scaling behavior of the 
synaptic weight configuration. We consider both a large range for the
number $N_w$ of afferent neurons and an extended range for the incoming 
signal-to-noise ratio. The input activity distributions $p(y_j)$ 
are Gaussians with standard deviations $\sigma_j=\sigma_\perp$ 
for ($j=2,\dots,N_w$), and with the dominant direction having a width
$\sigma_1$. We define the incoming signal-to-noise ratio as 
$S_i=\sigma_1/\sigma_\perp$, and investigate values for $S_i$ of 
2:1, 4:1, 8:1, 16:1 and 32:1. Shown in Fig.~\ref{fig_N_scaling} 
is the evolution of the outgoing signal-to-noise ratio, as a function of 
inputs $N_w$, and the evolution of the angle $\alpha$. All simulation
parameters are kept otherwise constant.

We define the outgoing signal-to-noise ratio as
$S_w=|w_1|/\sigma_w^{(non)}$ where $w_1$ is the synaptic weight
along the principal component and $\sigma_w^{(non)}$ the standard
deviation of the remaining synaptic weights (compare Eq.~(\ref{eq_alpha_N_w}) 
of the appendix). The outgoing signal-to-noise ratio is 
remarkably independent of the actual number $N_w$ of 
afferent neurons. $S_w$ shows, in addition, a threshold behavior, 
remaining finite even for data input streams characterized by 
small $S_i$. For large value of incoming signal-to-noise ratio 
a linear scaling $S_w\propto S_i$ is recovered.
 
Regarding the angle $\alpha$, the performance deteriorates, which 
increases steadily with $N_w$. This is however a dominantly statistical 
effect. In the appendix we show how the angle $\alpha$ increases
with $N_w$ for a constant outgoing signal-to-noise ratio $S_w$. 
This effect is then just a property of angles in large dimensional 
spaces and is independent of the learning rule employed.

\begin{figure}[!t]
\begin{center}
\includegraphics[width=0.6\textwidth]{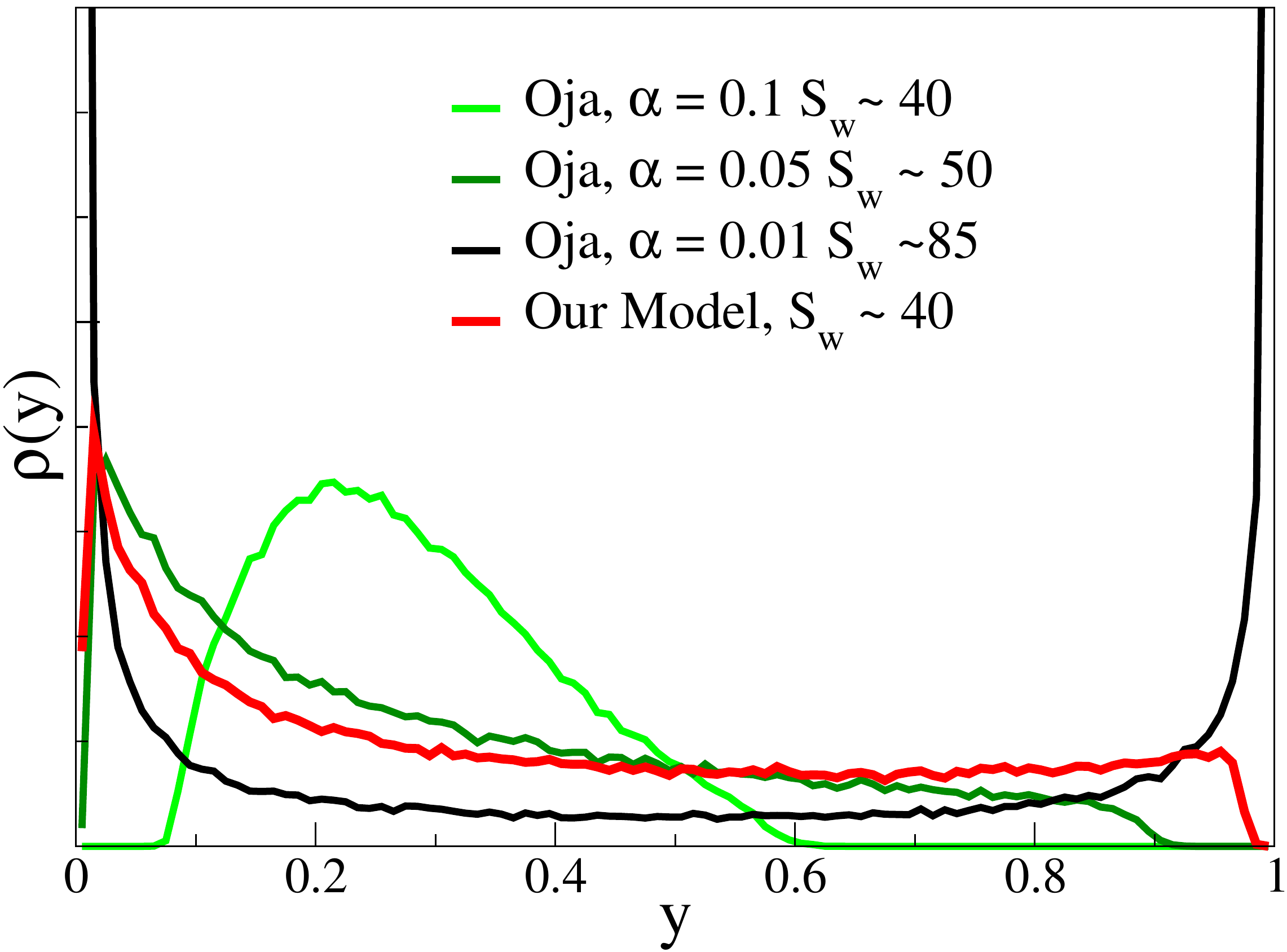}
\end{center}
\caption{
{The output distribution function and the signal to noise ratio.} 
The averaged firing-rate distributions $p(y)$ for $N_w=100$ and
and the parameter set used previously, compare Fig.~\ref{fig_N_scaling}.
In comparison the $p(y)$ resulting when using a modified Oja's rule,
see Eq.~(\ref{eq_Oja}), for the synaptic plasticity. Depending
on the parameter $\epsilon$, controlling the strength of the
weight decay in (\ref{eq_Oja}), arbitrary large signal-to-noise 
ratios $S_w$ can be achieved, on the expense of obtaining binary output
distributions. Note that the form of $p(y)$ is roughly comparable,
for similar signal-to-noise ratios, for the two approaches. The
output tends to cluster, however, around the target mean for
smaller $S_w$ and Oja's rule.
        }
\label{fig_distributions}
\end{figure}

It is interesting to compare the simulation results 
with other updating rules, like Oja's rule \citep{oja1997nonlinear},
\begin{equation}
\dot w_j \ =\ \epsilon_{oja} \left[y(y_j-\bar y_j) - 
\alpha\, y^2 w_j\right]~.
\label{eq_Oja}
\end{equation}
The original formulation used $\alpha=1$ for the relative
weighting of the decay term in (\ref{eq_Oja}). We find
however, for the case of non-linear neurons considered here,
that Oja's rule does not converge for $\alpha\gtrsim 0.1$.
For the results presented in Fig.~\ref{fig_distributions}
we adapted the bias using (\ref{eq_b_dot}) both
when using Oja's rule (\ref{eq_Oja}) and for our 
plasticity rule (\ref{eq_w_dot}). The parameter $\epsilon_{oja}$ 
was chosen such that the learning times (or the number of input 
patterns) needed for convergence matched, in this case 
$\epsilon_{oja}$ = 0.1. With Oja's rule, arbitrarily
large outgoing signal-to-noise ratios are achievable for 
$\alpha\to0$. In this case the resulting $p(y)$ becomes 
binary, as expected.  There is hence a trade-off and only 
intermediate values for the outgoing signal-to-noise ratio 
are achievable for smooth firing-rate distributions $p(y)$.
Note that Sanger's rule \citep{sanger1989optimal} reduces to
Oja's rule for the case of a single neuron, as considered here.


We also attempted to compare with the results of the BCM theory
\citep{bienenstock1982theory,intrator1992objective,cooper2012bcm}.
The BCM update rule also finds nicely the direction of
the principal component, but runaway synaptic growth occurs
generically in the case of the type of neurons considered in 
our study, being non-linear and  having an maximal possible 
firing rate, with $y\in[0,1]$. This is due to the fact 
that the upper cut-off of the firing rate preempts, in general, 
the sliding threshold to raise to values necessary to induce 
a large enough amount of synaptic weight decay. For the input 
distributions used throughout this study we could not 
avoid runaway synaptic growth for the BCM rule.

\subsection{Linear discrimination}

\begin{figure}[!t]
\begin{center}
\begin{tabular}{rr}
\raisebox{1ex}{\bf (A)}\includegraphics[width=0.32\textwidth]{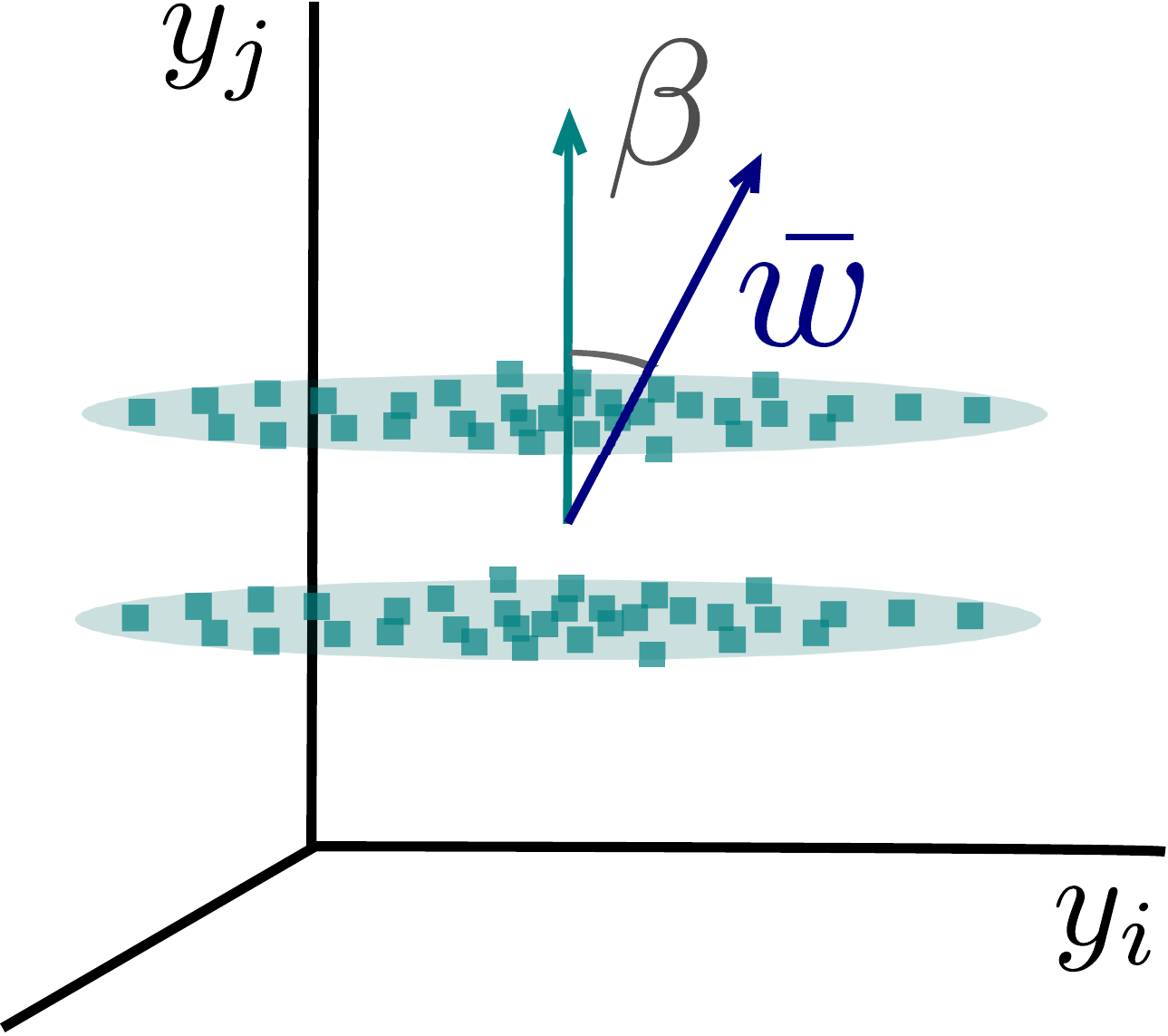}&
\raisebox{1ex}{\bf (B)}\includegraphics[width=0.37\textwidth]{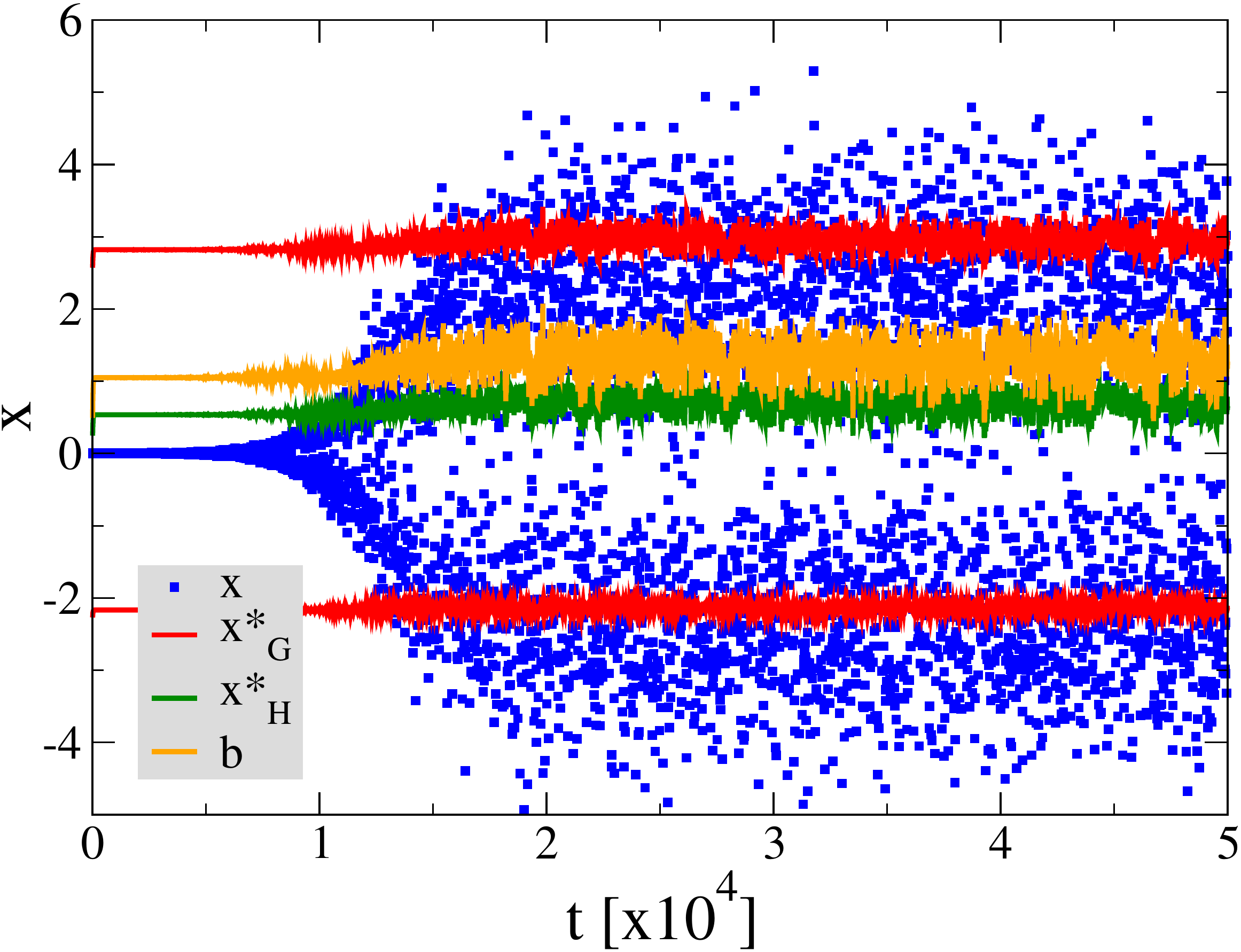}\\
\raisebox{1ex}{\bf (C)}\includegraphics[width=0.32\textwidth]{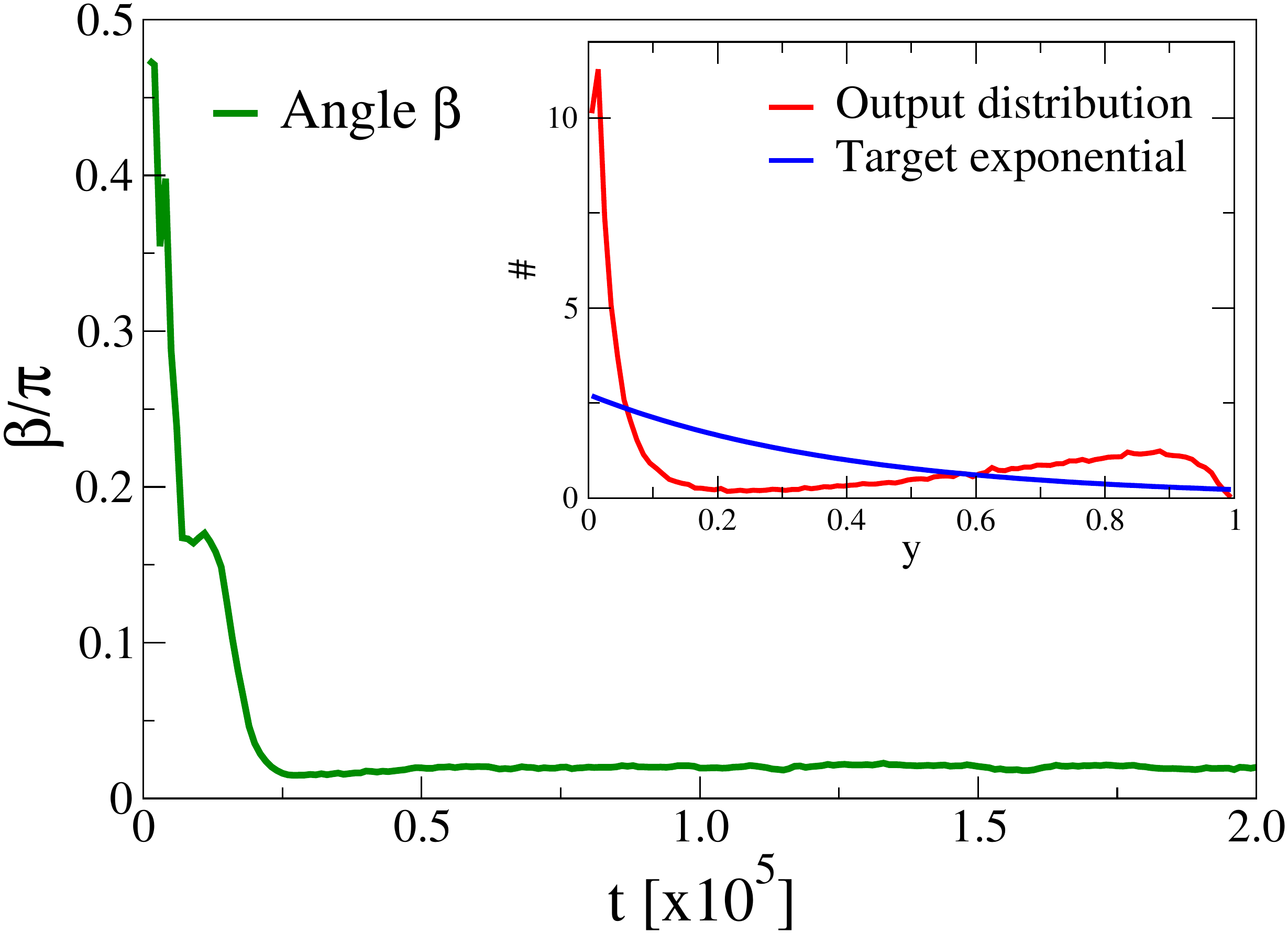} &
\raisebox{1ex}{\bf (D)}\includegraphics[width=0.37\textwidth]{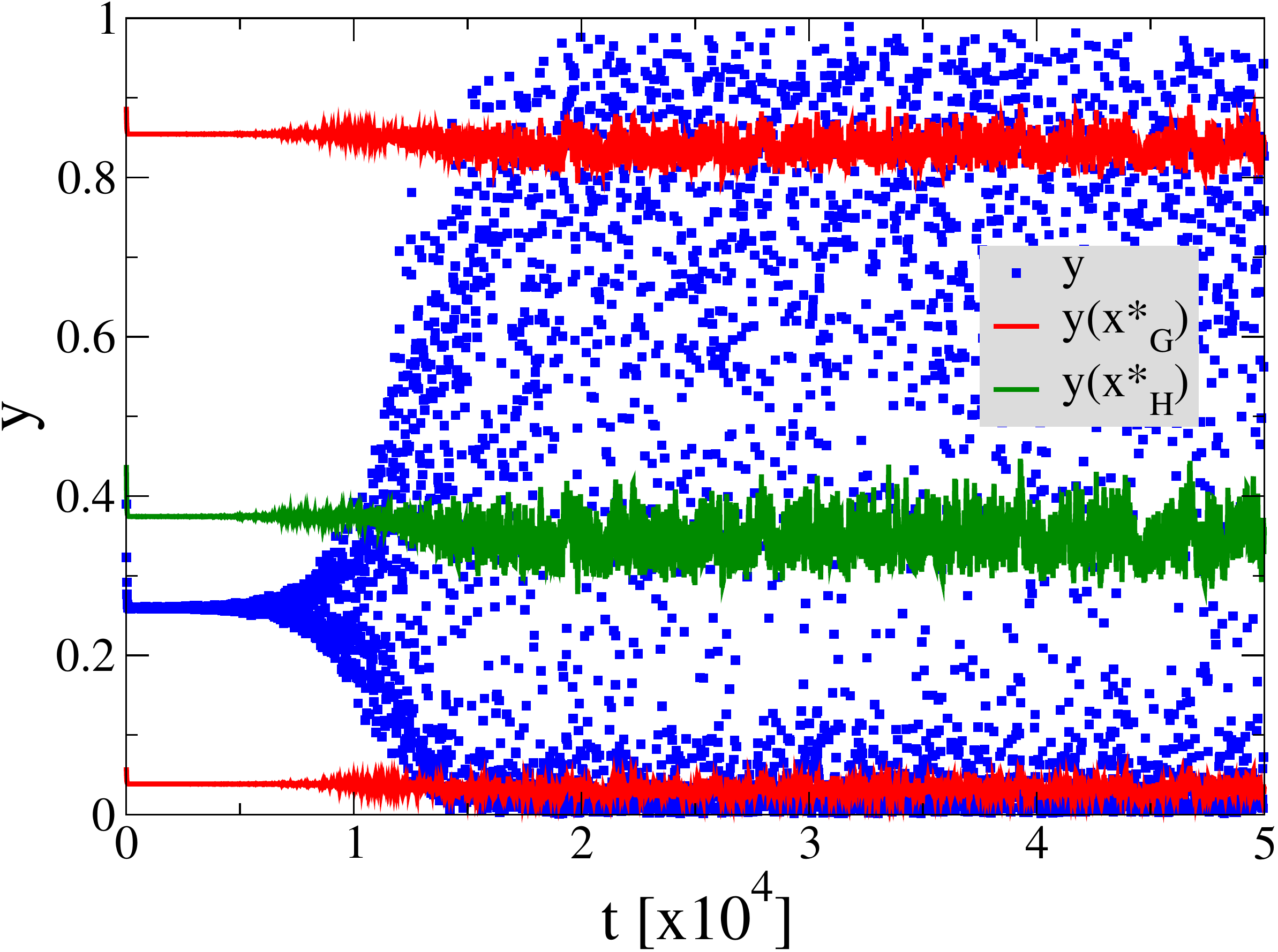}\\
\raisebox{1ex}{\bf (E)}\includegraphics[width=0.32\textwidth]{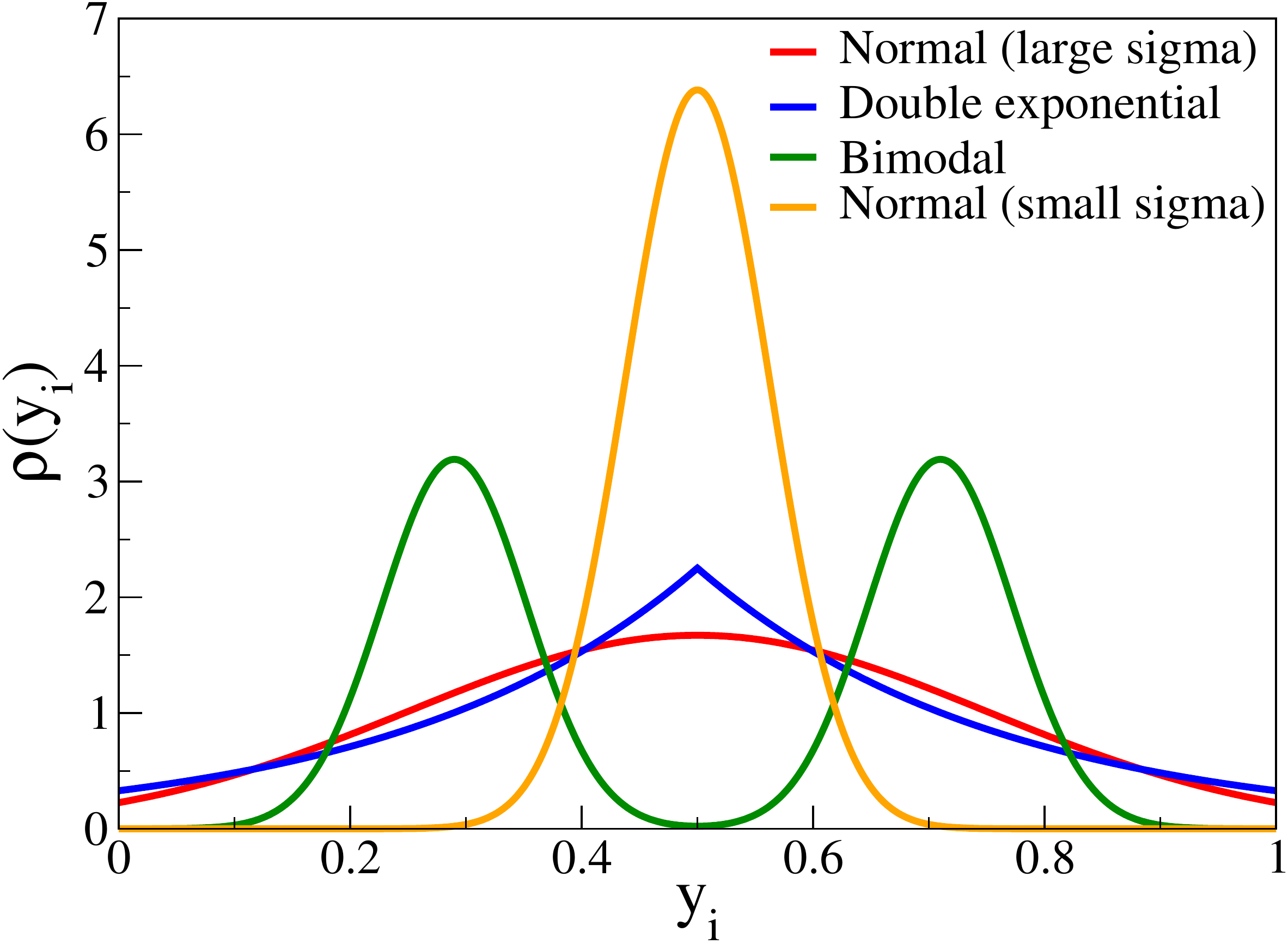}&
\raisebox{1ex}{\bf (F)}\includegraphics[width=0.37\textwidth]{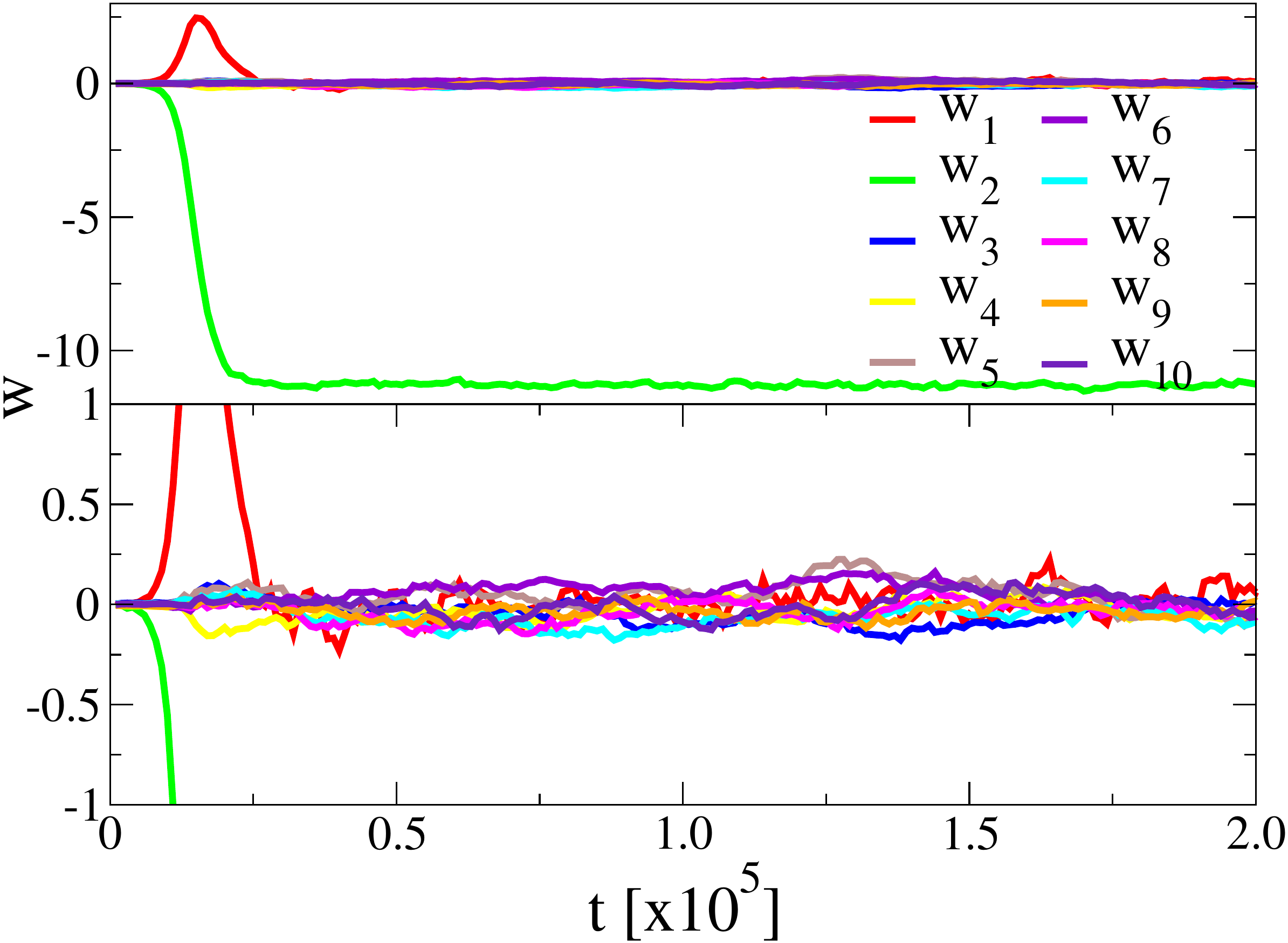}
\end{tabular}
\end{center}
\caption{
{Linear discrimination of bimodal input distributions.} 
Simulation results for a neuron with $N_w=100$ with
two directions having the same variance (but one being
bimodal) and the other $N_w-2$ directions
having a standard deviation four times smaller.
\textbf{(A)} 
Illustration of the input distribution density $p(y_1,y_2,\dots)$. 
\textbf{(B)} 
Time series of the membrane potential $x$ (blue), the bias $b$ (yellow),
the roots $x^*_G$ of the limiting factor $G(x)$ (red) and the
root $x^*_H$ of the Hebbian factor $H(x)$ (green).
\textbf{(C)} 
The evolution of the angle $\beta$ of the synaptic weight vector 
$\mathbf{w}$ with respect to the axis linking the two ellipsoids and 
(inset) the output distribution $p(y)$ (red) in comparison to the target
exponential (blue).
\textbf{(D)} 
Time series of the output $y$ (blue) and of the roots
$y^*_G$ of the limiting factor $G(y)$ (red) and the
root $y^*_H$ of the Hebbian factor $H(y)$ (green).
\textbf{(E)} Illustration of the distribution functions used,
the bimodal competing with the Normal distributed (alternatively
with a double exponential) having the same variance, all other
directions being normally distributed with a four times smaller
standard distribution.
\textbf{(F)} 
Time evolution of the first ten synaptic weights $\{w_j\}$,
separately for the principal component (upper panel) and
for nine other orthogonal directions (lower panel).
}
\label{fig_LDT}
\end{figure}

An important question regards the behavior of neural learning
rules when no distinct principal component is present in the
data input stream. In Fig.~\ref{fig_LDT} we present data for
the situation where two dominant directions have the same
standard deviation $\sigma\approx 0.22$, here for $p(y_1)$ 
and $p(y_2)$, with the remaining $N_w-2$ directions having 
a smaller standard deviation $\sigma/4$. In our experiment 
the first direction, $y_1$ is a unimodal Gaussian, as 
illustrated in Fig.~\ref{fig_LDT}(A), with the second 
direction, $y_2$ being bimodal. The two superposed Gaussian 
distributions along $y_2$ have individual widths $\sigma/4$ 
and the distance between the two maxima has been adjusted so 
that the overall standard deviation along $y_2$ is also $\sigma$. 


The synaptic weight vector aligns, for most randomly drawn 
starting ensembles $\{w_j\}$, with the bimodal direction, 
as shown in Fig.~\ref{fig_LDT}(C). In this case the system 
tries to adjust its parameters, namely the synaptic
weights and the bias $b$ so that the two peaks of the 
bimodal principal component are close to the two zeros $x^*_G$ 
(red symbols in Fig.~\ref{fig_LDT}(B))
of the limiting factor $G(x)$ in the adaption rule 
(\ref{eq_w_dot}). This effect is clearly present in the 
results for the membrane potential (blue symbols in 
Fig.~\ref{fig_LDT}(B)), clustering around the roots
of $G(x)$. The system performs, as a result, a linear 
discrimination with a bimodal output firing rate, 
presented in Fig.~\ref{fig_LDT}(D).

One possibility to characterize the deviation of a probability
distribution from a Gaussian is the excess kurtosis $\kappa$
\citep{decarlo1997meaning},
\begin{equation}
\kappa = \frac{Q_j}{\sigma_j^4}-3,
\quad\quad
Q_j = \int (y_j-\bar y_j)^4p(y_j)dy_j,
\quad\quad
\sigma_j^2 = \int (y_j-\bar y_j)^2p(y_j)dy_j~,
\label{eq_excess_kurtosis}
\end{equation}
with the normal distribution having, by construction, a
vanishing $\kappa\to0$. The excess kurtosis tends to be small
or negative on a finite support $p_j\in[0,1]$. Distributions characterized
by a positive $\kappa$ show pronounced tails. This statement
also holds for truncated Gaussians, as used in our simulations.
We have generalized the experiment presented in Fig.~\ref{fig_LDT} by 
studying the pairwise competition between three distributions
having all the same standard deviation $\sigma$, but varying
values of $\kappa$, compare Fig.~\ref{fig_LDT}(E): A
bimodal distribution with $\kappa =-1.69$, a unimodal Gaussian 
with $\kappa = -0.63$ and a unimodal double exponential
with $\kappa=-0.43$.

Running the simulation one thousand times, with randomly drawn
initial conditions, the direction with lower $\kappa$ was
selected $88.8\%$ / $65.4\%$ / $64.0\%$ of the times 
when the competing directions were bimodal vs.\ double exponential
/ Gaussian vs.\ double exponential / bimodal vs.\ Gaussian.
In none of the cases would both the first and the second synaptic
weights, $w_1$ and $w_2$, acquire large absolute
values. 

The underlying rationale for the updating rules favoring directions 
with negative excess kurtosis can be traced back to the inherent 
symmetry $F_{ob}(-x,1-y))=F_{ob}(x,y))$ of the objective function 
(\ref{eq_objectiveFunction}), which in turn is a consequence of treating 
both large and small firing rates on an equal footing in $F_{ob}$. There 
are two equivalent minima for $F_{ob}$ to which the maxima of a binary 
distribution are mapped, as discussed in section \ref{Theory}.

We have repeated this simulation using the modified Oja's rule 
(\ref{eq_Oja}), using $\alpha=0.1$ and $\epsilon_{oja}=0.1$.
We find a very distinct sensitivity, with the relative probability
for a certain input direction to be selected being
$97.0\%$ / $99.8\%$ / $42.1\%$ 
when the competing directions were bimodal vs.\ double exponential
/ Gaussian vs.\ double exponential / bimodal vs.\ Gaussian.
Note that all our input distributions are centered around 0.5
and truncated to $[0,1]$. Oja's rule has a preference for
unimodal distributions and a strong dislike of double 
exponentials. However, the excess kurtosis does not seem to
be a determining parameter, within Oja's rule, for the 
directional selectivity.

\subsection{Continuous online learning - fading memory}

Another aspect of relevance concerns the behavior of synaptic plasticity 
rules for continuous online learning. A basic requirement is the absence 
of runaway growth effects in the presence of stationary input statistics.
But how should a neuron react when the statistics of the afferent input 
stream changes at a certain point? Should it adapt immediately, at a very 
short time scale or should it show a certain resilience, adapting to the 
new stimuli only when these show a certain persistence?

\begin{figure}[!t]
\begin{center}
\includegraphics[width=0.65\textwidth]{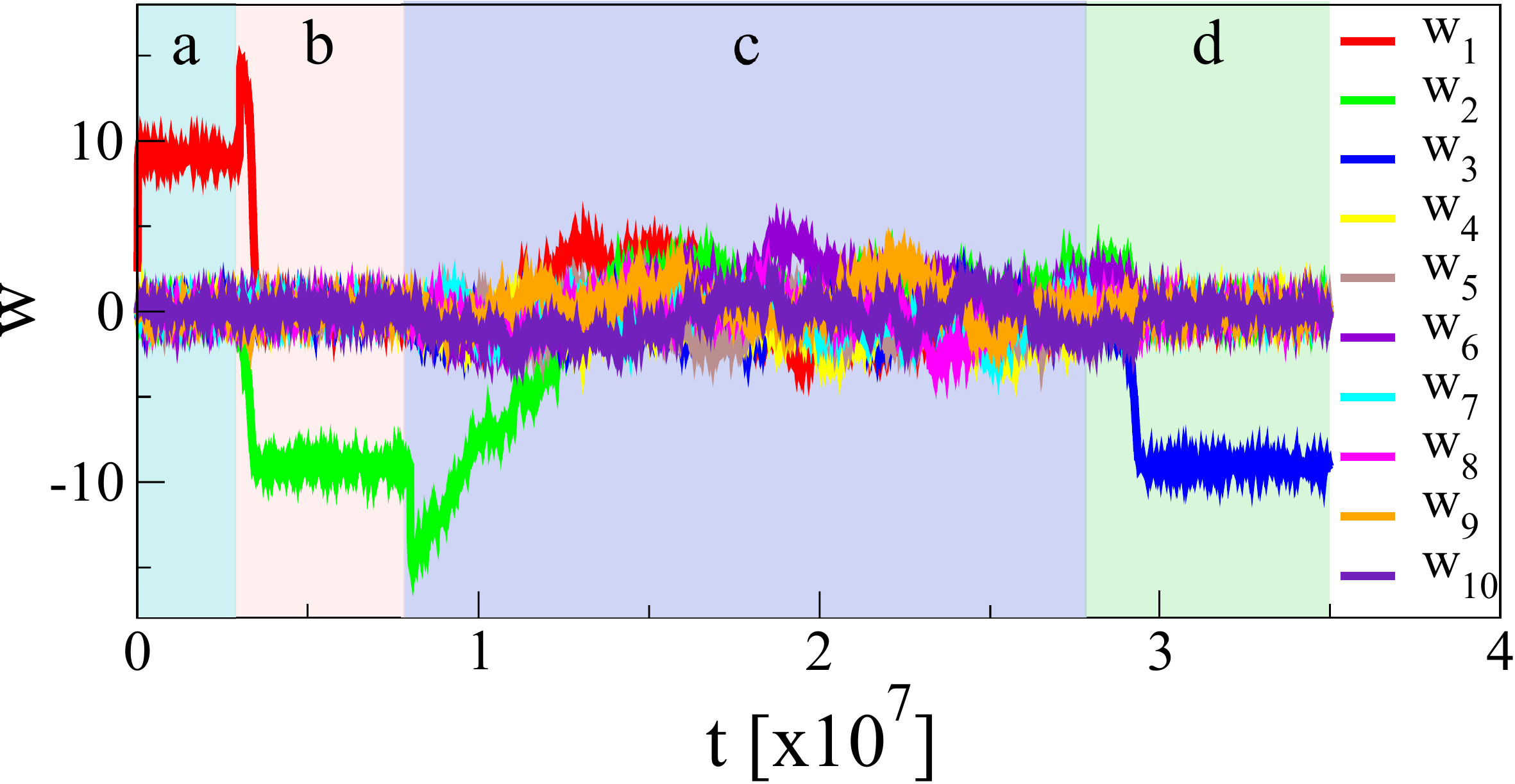}
\vspace{2ex}

\includegraphics[width=0.65\textwidth]{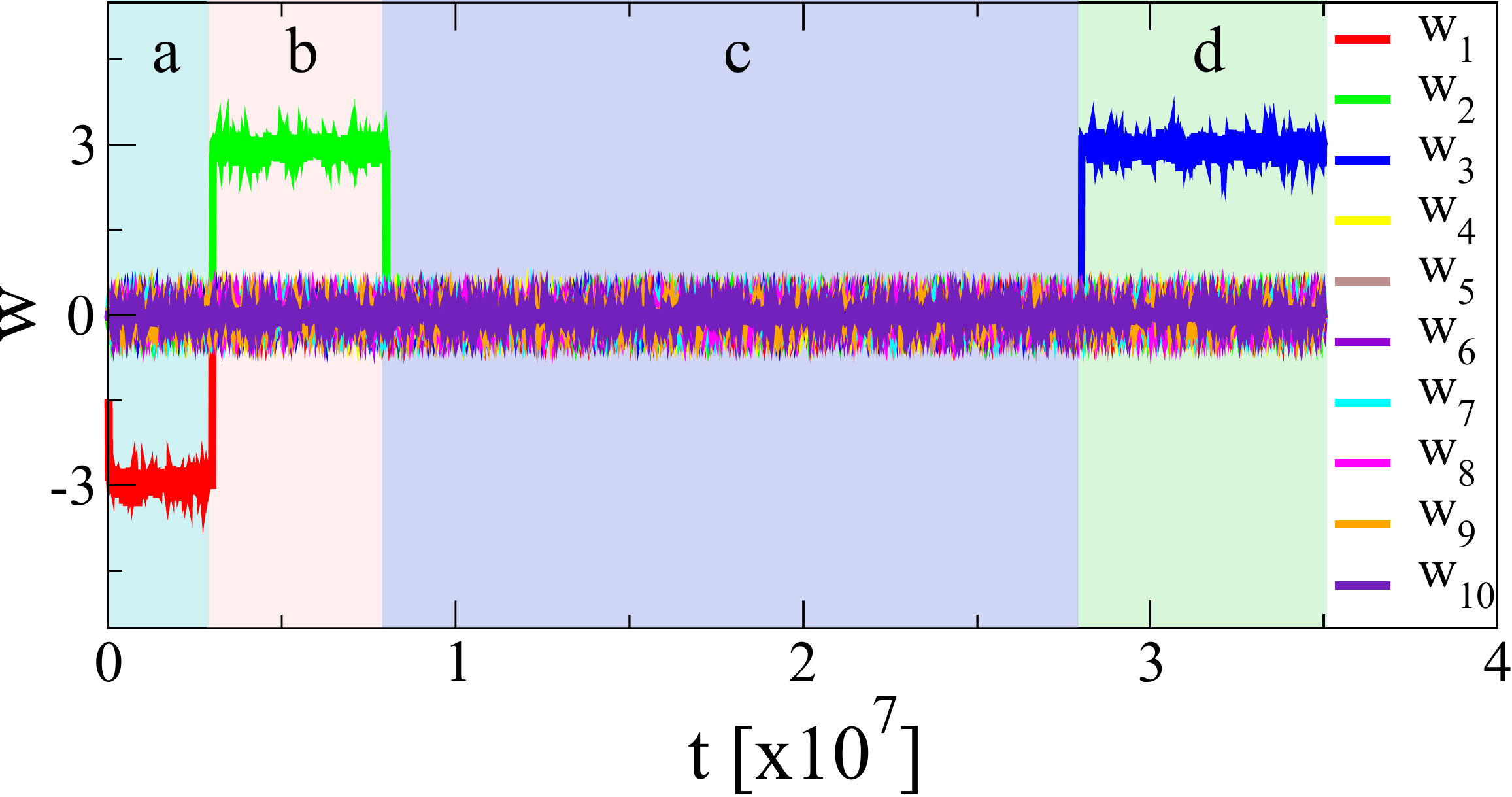}
\end{center}
\caption{
{Continuous online learning and weak forgetting.} 
The effect of changing the statistics of the input firing
rates $p(y_j)$. During (a), (b) and (d) the principal axis
is along $y_1$, $y_2$ and $y_3$ respectively, during (c)
there in no principal component. There are $N_w=100$ 
afferent neurons, shown is the time evolution of the first 
ten synaptic weights. The standard deviations of the afferent 
neurons is $\sigma$ for the principal direction, if there
is any, and $\sigma/2$ for all other directions,
compare Fig.~\ref{fig_PCA}. During(c) all inputs have
identical standard deviations $\sigma/2$. The initial weight 
distribution is randomly drawn. The top/bottom panel
show the results using respectively our synaptic updating 
rule (\ref{eq_w_dot}) and Oja's rule (\ref{eq_Oja}).
Note that it takes considerably longer, for our updating rule,
to unlearn than to learn from scratch. Learning and unlearning
occurs, on the other side, at the same timescale for Oja's rule.
        }
\label{fig_online_learning}
\end{figure}

We have examined the behavior of the adaption rules upon a sudden change 
of firing-rate statistics of the afferent neurons. We find, as presented 
in Fig.~\ref{fig_online_learning}, that the new statistics is recognized 
autonomously, with a considerable resilience to unlearn the previously 
acquired information about the statistics of the input data stream. The 
synaptic plasticity rules (\ref{eq_w_dot}) do hence incorporate a fading 
memory.


In our experiment we considered $N_w=100$ afferent neurons, with Gaussian 
firing distributions having standard deviation $\sigma$ for the principal 
component and $\sigma/2$ for the remaining $N_w-1$ directions. The sign 
of the synaptic weights are not of relevance, as the input distributions 
$p(y_j)$ are symmetric with respect to their means, taken to be $0.5$. 
The direction of the principal component is then changed several times, 
everything else remaining otherwise unchanged. 

The starting configuration 
$\{w_j\}$ of synaptic weights has been drawn randomly from $[-0.005:0005]$ 
and the initial learning is fast, occurring on a time scale of 
$T_{initial}\approx 10^4$ updatings, compare Fig.~\ref{fig_PCA}, using the 
same updating rates $\epsilon_w=0.01$ and $\epsilon_b=0.1$ as throughout 
this paper. The time the neuron takes to adapt to the new input statistics 
is however of the order of $T_{unlearn}\approx 10^6$, viz about two orders 
of magnitude larger than $T_{initial}$. New information is hence acquired 
at a slower rate; the system shows a substantial resilience to unlearn 
previously acquired memories. 

One can observe in Fig.~\ref{fig_online_learning} an overshoot of the
principal synaptic weight, just before the unlearning starts, as
the system tries to keep the  membrane potential $x$ within its
working regime, compare Fig.~\ref{fig_PCA}(D). The system reacts
by increasing the largest synaptic weight when the variance
of the input drops along the corresponding afferent direction, before
it can notice that the principal component of the afferent activities
has also changed.

Also included in the simulation presented in Fig.~\ref{fig_online_learning} 
is a phase without any principal component, the statistics of all incoming 
$p(y_j)$ being identical, viz with the covariance matrix being proportional 
to unity. One notices that the neuron shows a marked resilience to forget the 
previously acquired knowledge, taking about $5\cdot 10^7$ updates in order to 
return to a fully randomized drifting configuration of synaptic weights 
$\{w_j\}$. The synaptic plasticity rule (\ref{eq_w_dot}) hence leads to an
extended fading memory, which we believe to be a consequence of its
multiplicative structure.

For comparison we have repeated the same experiment using the modified 
Oja's rule (\ref{eq_Oja}), using $\alpha=0.1$ (which yields the same 
signal-to-noise ratio, compare Fig.~\ref{fig_distributions}), and
$\epsilon_{oja}=0.1$, such that the initial learning rates 
(achieving 90\% of the stationary value for the principal component) 
are comparable for both updating rules. We also kept the same updating 
(\ref{eq_b_dot}) for the bias. For Oja's rule learning and unlearning occurs 
on very similar time scales, reacting immediately to changes in the statistics 
of the input activities.

It is presently not entirely clear which form of unlearning is present in the 
brain, on the level of individual neurons. While studies in prefrontal cortex have 
shown full learning and unlearning of different categories in binary 
classification tasks, related in this context to the concept of adaptive coding 
\citep{duncan2001adaptive}, more complex behavioral responses tend 
however to exhibit slow or incomplete unlearning such as extinction of paired 
cue - response associations, in the context of Pavlovian conditioning 
\citep{myers2002behavioral,quirk2007neural}. It is also conceivable that a 
fading memory may possibly be advantageous in the context of noisy environments
with fluctuating activity statistics.

\section{Discussion}
\label{Discussion}

Objective functions based on information theoretical principles
play an important role in neuroscience 
\citep{intrator1992objective,lengelle1996training,goodhill1997unifying,kay2011coherent}
and cognitive robotics \citep{sporns2006evolving,ay2008predictive}.
Many objective functions investigated hitherto use either 
Shannon's information directly, or indirectly by considering
related measures, like predictive and mutual entropy
\citep{kraskov2004estimating}, or the Kullback-Leibler 
divergence. Objective functions are instances,
from a somewhat larger perspective, of generating functionals, as
they are normally used to derive equations of motion for the neural
activity, or to deduce adaption rules for secondary variables 
like synaptic weights or intrinsic parameters. Here we discuss 
an objective function which may be either motivated by its 
own virtue, as discussed in section \ref{Theory}, or 
by considering the Fisher information as a generating functional.

The Fisher information encodes the sensitivity of a given
probability distribution function, in our case the distribution
of neural firing rates, with respect to a certain parameter
of interest. Cognitive information processing in the brain is 
all about changing the neural firing statistics and we hence
believe that the Fisher information constitutes an interesting 
starting point from where to formulate guiding principles for 
plasticity in the brain or in artificial systems. In particular, 
we have examined the Fisher information with respect to changes 
of the synaptic weights. Minimizing this objective function, 
which we denoted as the synaptic flux, we find self-limiting 
adaption rules for unsupervised and autonomous learning. The 
adaption rules are Hebbian, with the self limitation leading 
to synaptic competition and an alignment of the synaptic weight 
vector with the principal component of the input data stream. 

Synaptic plasticity rules for rate encoding neurons are crucial
for artificial neural networks used for cognitive tasks and 
machine learning, and important for the interpretation of the 
time-averaged behavior of spiking neurons. In this context our 
adaption rules make two predictions, which one may eventually test 
experimentally. The first prediction concerns the adaption in the 
situation where more than one dominant component is present in the 
space of input activities. Our model implies for this case a robust 
tendency for the synaptic weight vector to favor directions in the space 
of input activities being bimodal, characterized by a negative kurtosis.

Our adaption rules have a second implication, regarding the robustness 
of acquired memories with respect to persistent changes of the statistics 
of the input activities, in the context of continuous and unsupervised
online learning. We predict that it is considerably easier for the 
neuron to detect relevant features in the space of input activities 
when starting from a virgin state of a random synaptic configuration. 
New features will still be extracted from the stream of input activities,
and old ones unlearned at the same time, once the initial synaptic adaption 
process has been completed, albeit at a much slower pace. This feature can 
be interpreted as a sturdy fading memory.

We have extensively examined the robustness of the behavior of 
the synaptic plasticity rules upon variation of the simulation setup. 
All results presented here remain fully valid when changing, e.g.\ 
the adaption rate $\epsilon_w$, in particular we have examined 
$\epsilon_w=0.1$ and $\epsilon_w=0.001$. We have also studied other 
forms of input activities $p(y_j)$ and found only quantitative changes 
for the response. For example, we have considered exponentially 
distributed input statistics, as a consistency check with the target 
output distribution function. We hence believe that the here proposed 
synaptic plasticity rules are robust to a considerable degree, a 
prerequisite for viable plasticity rules, both in the context of biological 
and artificial systems.

The synaptic plasticity rule (\ref{eq_w_dot}) is a product of 
two conjugate factors, the limiting factor $G(x)$ and the 
Hebbian factor $H(x)$. Runaway synaptic growth occurs,
as we have verified numerically, when setting $G(x)$ to a constant.
Unlimited synaptic growth occurs despite the emergence of a sliding
threshold (see Eq.~(\ref{eq_H_y}) of section \ref{Theory}) as
the firing rate $y(t)\in[0,1]$ is bounded. Runaway synaptic growth
results in increasing (positive and negative) large membrane potentials
$x(t)$, with the firing becoming binary, accumulating at the boundaries,
viz $y\to0$ and $y\to1$. 

Finally we comment on the conceptual foundations of this work. 
The adaptive time evolution of neural networks and the continuous 
reconfiguration of synaptic weights may be viewed as a self-organizing 
processes guided by certain target objectives 
\citep{gros2010complex,prokopenko2009guided,friston2010free,linkerhand2013generating}.
A single objective function will in general not be enough for 
generating dynamics of sufficient complexity, as necessary for neural 
circuitry or synaptic reconfiguration processes. It has indeed been 
noted that the interplay between two or more generating functionals 
may give rise to highly nontrivial dynamical states 
\citep{linkerhand2013generating,gros2014generating}. 

In this context, it is important to note that several generating 
functionals may in general not be combined to a single overarching 
objective function. Dynamical systems can hence show, under the
influence of competing objective functions, complex self-organizing
behavior \citep{linkerhand2013generating,gros2014generating}. 
In the present work we propose that the interplay between two specific
objective functions, namely the Fisher information for the synaptic 
flux and the Kullback-Leibler divergence for the information content 
of the neural firing rate, give rise, quite naturally, to a set of 
viable adaption rules for self-limiting synaptic and intrinsic plasticity 
rules.

\section{Acknowledgments}
\label{Acknowledgments}

R.E.\ acknowledges stimulating discussions at the
OCCAM 2013 workshop. The support of the German Science
Foundation (DFG) is acknowledged.

\section{Appendix: Modeling adaption for large numbers of 
transversal directions}
\label{Large_N}

For simulations with $N_w$ Gaussian input distributions $p(y_j)$
the synaptic weight vector adapts to
\begin{equation}
\mathbf{w} = (w_1,w_2,\dots,w_{N_w}),
\qquad\quad  w_1\gg w_k\qquad (k\ge2)~,
\label{eq_PCA_w}
\end{equation}
when $p(y_1)$ is assumed to have the largest standard deviation 
$\sigma_1$, with all other $p(y_k)$, for $k=2,\dots, N_w$  
having a smaller standard deviation $\sigma_k$. The angle 
$\alpha$ between the synaptic weight vector and the direction 
$(1,0,\dots,0)$ of the principal component is hence given by
\begin{equation}
\cos(\alpha) = \frac{w_1}{\sqrt{w_1^2+\sum_{k>1} w_k^2}}
             = \frac{w_1}{\sqrt{w_1^2+(N_w-1)\left(\sigma_w^{(non)}\right)^2}}
\approx \frac{1}{\sqrt{N_w}} \frac{w_1}{\sigma_w^{(non)}} ~,
\label{eq_alpha_N_w}
\end{equation}
where we have defined with 
$\sigma_w^{(non)}=\left(\sum_{k>1} w_k^2\right)/(N_w-1)$ 
the averaged standard deviation of the non-principal components 
(which have generically a vanishing mean). In our simulation
we find, compare Fig.~\ref{fig_N_scaling}, an outgoing signal-to-noise 
ratio $S_w=|w_1|/\sigma_w^{(non)}$ which is remarkably 
independent of $N_w$ and hence that $\alpha$ approaches 
$\pi/2$ like $\pi/2-r/\sqrt{N_w}$ in the limit of large 
numbers $N_w\to\infty$ of afferent neurons, where $r$
is a constant, independent of $N_w$. This statistical
degradation of the performance, in terms of the angle $\alpha$,
is hence a variant of the well known curse of dimensionality
\citep{jain2000statistical}.



\bibliographystyle{frontiersinSCNS&ENG}
\bibliography{synapticFlux-frontiers}

\begin{thebibliography}{52}
\providecommand{\natexlab}[1]{#1}
\expandafter\ifx\csname urlstyle\endcsname\relax
  \providecommand{\doi}[1]{doi:\discretionary{}{}{}#1}\else
  \providecommand{\doi}{doi:\discretionary{}{}{}\begingroup
  \urlstyle{rm}\Url}\fi
\providecommand{\selectlanguage}[1]{\relax}

\bibitem[{\textbf{Abbott and Nelson}(2000)}]{abbott2000synaptic}
Abbott, L.~F. and Nelson, S.~B. (2000), Synaptic plasticity: taming the beast,
  \emph{Nature Neuroscience}, 3, 1178--1183

\bibitem[{\textbf{Ay et~al.}(2008)\textbf{Ay, Bertschinger, Der, G{\"u}ttler,
  and Olbrich}}]{ay2008predictive}
Ay, N., Bertschinger, N., Der, R., G{\"u}ttler, F., and Olbrich, E. (2008),
  Predictive information and explorative behavior of autonomous robots,
  \emph{The European Physical Journal B}, 63, 3, 329--339

\bibitem[{\textbf{Bell and Sejnowski}(1995)}]{bell1995information}
Bell, A.~J. and Sejnowski, T.~J. (1995), An information-maximization approach
  to blind separation and blind deconvolution, \emph{Neural Computation}, 7, 6,
  1129--1159

\bibitem[{\textbf{Bienenstock et~al.}(1982)\textbf{Bienenstock, Cooper, and
  Munro}}]{bienenstock1982theory}
Bienenstock, E.~L., Cooper, L.~N., and Munro, P.~W. (1982), Theory for the
  development of neuron selectivity: orientation specificity and binocular
  interaction in visual cortex, \emph{The Journal of Neuroscience}, 2, 1,
  32--48

\bibitem[{\textbf{Brunel and Nadal}(1998)}]{brunel1998mutual}
Brunel, N. and Nadal, J.-P. (1998), Mutual information, fisher information, and
  population coding, \emph{Neural Computation}, 10, 7, 1731--1757

\bibitem[{\textbf{Chagnac-Amitai et~al.}(1990)\textbf{Chagnac-Amitai, Luhmann,
  and Prince}}]{chagnac1990bursts}
Chagnac-Amitai, Y., Luhmann, H.~J., and Prince, D.~A. (1990), Burst generating
  and regular spiking layer 5 pyramidal neurons of rat neocortex have different
  morphological features, \emph{Journal of Comparative Neurology}, 296, 4,
  598--613

\bibitem[{\textbf{Cooper and Bear}(2012)}]{cooper2012bcm}
Cooper, L.~N. and Bear, M.~F. (2012), The bcm theory of synapse modification at
  30: interaction of theory with experiment, \emph{Nature Reviews
  Neuroscience}, 13, 11, 798--810

\bibitem[{\textbf{DeCarlo}(1997)}]{decarlo1997meaning}
DeCarlo, L.~T. (1997), On the meaning and use of kurtosis, \emph{Psychological
  Methods}, 2, 3, 292--307

\bibitem[{\textbf{DiCarlo et~al.}(2012)\textbf{DiCarlo, Zoccolan, and
  Rust}}]{dicarlo2012does}
DiCarlo, J.~J., Zoccolan, D., and Rust, N.~C. (2012), How does the brain solve
  visual object recognition?, \emph{Neuron}, 73, 3, 415--434

\bibitem[{\textbf{Dong and Hopfield}(1992)}]{dong1992dynamic}
Dong, D.~W. and Hopfield, J.~J. (1992), Dynamic properties of neural networks
  with adapting synapses, \emph{Network: Computation in Neural Systems}, 3, 3,
  267--283

\bibitem[{\textbf{Duncan}(2001)}]{duncan2001adaptive}
Duncan, J. (2001), An adaptive coding model of neural function in prefrontal
  cortex, \emph{Nature Reviews Neuroscience}, 2, 11, 820--829

\bibitem[{\textbf{Elliott}(2003)}]{elliott2003analysis}
Elliott, T. (2003), An analysis of synaptic normalization in a general class of
  hebbian models, \emph{Neural Computation}, 15, 4, 937--963

\bibitem[{\textbf{Feldman}(2012)}]{feldman2012spike}
Feldman, D.~E. (2012), The spike-timing dependence of plasticity,
  \emph{Neuron}, 75, 4, 556--571

\bibitem[{\textbf{Friston}(2010)}]{friston2010free}
Friston, K. (2010), The free-energy principle: a unified brain theory?,
  \emph{Nature Reviews Neuroscience}, 11, 2, 127--138

\bibitem[{\textbf{Goodhill and Barrow}(1994)}]{goodhill1994role}
Goodhill, G.~J. and Barrow, H.~G. (1994), The role of weight normalization in
  competitive learning, \emph{Neural Computation}, 6, 2, 255--269

\bibitem[{\textbf{Goodhill and Sejnowski}(1997)}]{goodhill1997unifying}
Goodhill, G.~J. and Sejnowski, T.~J. (1997), A unifying objective function for
  topographic mappings, \emph{Neural Computation}, 9, 6, 1291--1303

\bibitem[{\textbf{Gros}(2010)}]{gros2010complex}
Gros, C. (2010), Complex and adaptive dynamical systems: A primer (Springer
  Verlag)

\bibitem[{\textbf{Gros}(2014)}]{gros2014generating}
Gros, C. (2014), Generating functionals for guided self-organization, in
  M.~Prokopenko, ed., Guided Self-Organization: Inception (Springer), 53--66

\bibitem[{\textbf{Gutnisky and Dragoi}(2008)}]{gutnisky2008adaptive}
Gutnisky, D.~A. and Dragoi, V. (2008), Adaptive coding of visual information in
  neural populations, \emph{Nature}, 452, 7184, 220--224

\bibitem[{\textbf{Hebb}(2002)}]{hebb2002organization}
Hebb, D.~O. (2002), The organization of behavior: A neuropsychological theory
  (Psychology Press)

\bibitem[{\textbf{Huber}(1985)}]{huber1985projection}
Huber, P.~J. (1985), Projection pursuit, \emph{The annals of Statistics},
  435--475

\bibitem[{\textbf{Intrator and Cooper}(1992)}]{intrator1992objective}
Intrator, N. and Cooper, L.~N. (1992), Objective function formulation of the
  bcm theory of visual cortical plasticity: Statistical connections, stability
  conditions, \emph{Neural Networks}, 5, 1, 3--17

\bibitem[{\textbf{Jain et~al.}(2000)\textbf{Jain, Duin, and
  Mao}}]{jain2000statistical}
Jain, A.~K., Duin, R. P.~W., and Mao, J. (2000), Statistical pattern
  recognition: A review, \emph{Pattern Analysis and Machine Intelligence, IEEE
  Transactions on}, 22, 1, 4--37

\bibitem[{\textbf{Kay and Phillips}(2011)}]{kay2011coherent}
Kay, J.~W. and Phillips, W. (2011), Coherent infomax as a computational goal
  for neural systems, \emph{Bulletin of mathematical biology}, 73, 2, 344--372

\bibitem[{\textbf{Kraskov et~al.}(2004)\textbf{Kraskov, St{\"o}gbauer, and
  Grassberger}}]{kraskov2004estimating}
Kraskov, A., St{\"o}gbauer, H., and Grassberger, P. (2004), Estimating mutual
  information, \emph{Physical review E}, 69, 6, 066138

\bibitem[{\textbf{Lengell{\'e} and Denoeux}(1996)}]{lengelle1996training}
Lengell{\'e}, R. and Denoeux, T. (1996), Training mlps layer by layer using an
  objective function for internal representations, \emph{Neural Networks}, 9,
  1, 83--97

\bibitem[{\textbf{Linkerhand and
  Gros}(2013{\natexlab{a}})}]{linkerhand2013generating}
Linkerhand, M. and Gros, C. (2013{\natexlab{a}}), Generating functionals for
  autonomous latching dynamics in attractor relict networks, \emph{Scientific
  Reports (in press)}

\bibitem[{\textbf{Linkerhand and
  Gros}(2013{\natexlab{b}})}]{linkerhand2013self}
Linkerhand, M. and Gros, C. (2013{\natexlab{b}}), Self-organized stochastic
  tipping in slow-fast dynamical systems, \emph{Mathematics and Mechanics of
  Complex Systems}, 1-2, 129

\bibitem[{\textbf{Lisman and Spruston}(2010)}]{lisman2010questions}
Lisman, J. and Spruston, N. (2010), Questions about stdp as a general model of
  synaptic plasticity, \emph{Frontiers in Synaptic Neuroscience}, 2

\bibitem[{\textbf{Lisman}(1997)}]{lisman1997bursts}
Lisman, J.~E. (1997), Bursts as a unit of neural information: making unreliable
  synapses reliable, \emph{Trends in neurosciences}, 20, 1, 38--43

\bibitem[{\textbf{Markovi{\'c} and Gros}(2010)}]{markovic2010}
Markovi{\'c}, D. and Gros, C. (2010), Self-organized chaos through
  polyhomeostatic optimization, \emph{Physical Review Letters}, 105, 6, 068702

\bibitem[{\textbf{Markovi{\'c} and Gros}(2012)}]{markovic2012intrinsic}
Markovi{\'c}, D. and Gros, C. (2012), Intrinsic adaptation in autonomous
  recurrent neural networks, \emph{Neural Computation}, 24, 2, 523--540

\bibitem[{\textbf{Miller and MacKay}(1994)}]{miller1994role}
Miller, K.~D. and MacKay, D.~J. (1994), The role of constraints in hebbian
  learning, \emph{Neural Computation}, 6, 1, 100--126

\bibitem[{\textbf{Myers and Davis}(2002)}]{myers2002behavioral}
Myers, K.~M. and Davis, M. (2002), Behavioral and neural analysis of
  extinction, \emph{Neuron}, 36, 4, 567--584

\bibitem[{\textbf{Nagy}(2003)}]{nagy2003fisher}
Nagy, A. (2003), Fisher information in density functional theory, \emph{The
  Journal of chemical physics}, 119, 18, 9401--9405

\bibitem[{\textbf{Oja}(1992)}]{oja1992principal}
Oja, E. (1992), Principal components, minor components, and linear neural
  networks, \emph{Neural Networks}, 5, 6, 927--935

\bibitem[{\textbf{Oja}(1997)}]{oja1997nonlinear}
Oja, E. (1997), The nonlinear pca learning rule in independent component
  analysis, \emph{Neurocomputing}, 17, 1, 25--45

\bibitem[{\textbf{Paradiso}(1988)}]{paradiso1988theory}
Paradiso, M. (1988), A theory for the use of visual orientation information
  which exploits the columnar structure of striate cortex, \emph{Biological
  cybernetics}, 58, 1, 35--49

\bibitem[{\textbf{Prokopenko}(2009)}]{prokopenko2009guided}
Prokopenko, M. (2009), Guided self-organization, \emph{HFSP Journal}, 3,
  287--289

\bibitem[{\textbf{Prokopenko et~al.}(2011)\textbf{Prokopenko, Lizier, Obst, and
  Wang}}]{prokopenko2011relating}
Prokopenko, M., Lizier, J.~T., Obst, O., and Wang, X.~R. (2011), Relating
  fisher information to order parameters, \emph{Physical Review E}, 84, 4,
  041116

\bibitem[{\textbf{Quirk and Mueller}(2007)}]{quirk2007neural}
Quirk, G.~J. and Mueller, D. (2007), Neural mechanisms of extinction learning
  and retrieval, \emph{Neuropsychopharmacology}, 33, 1, 56--72

\bibitem[{\textbf{Reginatto}(1998)}]{reginatto1998derivation}
Reginatto, M. (1998), Derivation of the equations of nonrelativistic quantum
  mechanics using the principle of minimum fisher information, \emph{Physical
  Review A}, 58, 1775--1778

\bibitem[{\textbf{Sanger}(1989)}]{sanger1989optimal}
Sanger, T.~D. (1989), Optimal unsupervised learning in a single-layer linear
  feedforward neural network, \emph{Neural networks}, 2, 6, 459--473

\bibitem[{\textbf{Seung and Sompolinsky}(1993)}]{seung1993simple}
Seung, H. and Sompolinsky, H. (1993), Simple models for reading neuronal
  population codes, \emph{Proceedings of the National Academy of Sciences}, 90,
  22, 10749--10753

\bibitem[{\textbf{Shouval et~al.}(2010)\textbf{Shouval, Wang, and
  Wittenberg}}]{shouval2010spike}
Shouval, H.~Z., Wang, S. S.-H., and Wittenberg, G.~M. (2010), Spike timing
  dependent plasticity: a consequence of more fundamental learning rules,
  \emph{Frontiers in Computational Neuroscience}, 4

\bibitem[{\textbf{Simoncelli and Olshausen}(2001)}]{simoncelli2001natural}
Simoncelli, E.~P. and Olshausen, B.~A. (2001), Natural image statistics and
  neural representation, \emph{Annual review of neuroscience}, 24, 1,
  1193--1216

\bibitem[{\textbf{Sinz and Bethge}(2013)}]{sinz2013temporal}
Sinz, F. and Bethge, M. (2013), Temporal adaptation enhances efficient contrast
  gain control on natural images, \emph{PLoS computational biology}, 9, 1,
  e1002889

\bibitem[{\textbf{Sporns and Lungarella}(2006)}]{sporns2006evolving}
Sporns, O. and Lungarella, M. (2006), Evolving coordinated behavior by
  maximizing information structure, in Artificial life X: proceedings of the
  tenth international conference on the simulation and synthesis of living
  systems, 323--329

\bibitem[{\textbf{Triesch}(2007)}]{triesch2007synergies}
Triesch, J. (2007), Synergies between intrinsic and synaptic plasticity
  mechanisms, \emph{Neural Computation}, 19, 4, 885--909

\bibitem[{\textbf{Turrigiano and Nelson}(2000)}]{turrigiano2000hebb}
Turrigiano, G.~G. and Nelson, S.~B. (2000), Hebb and homeostasis in neuronal
  plasticity, \emph{Current opinion in neurobiology}, 10, 3, 358--364

\bibitem[{\textbf{Vicente et~al.}(2011)\textbf{Vicente, Wibral, Lindner, and
  Pipa}}]{vicente2011transfer}
Vicente, R., Wibral, M., Lindner, M., and Pipa, G. (2011), Transfer entropy???a
  model-free measure of effective connectivity for the neurosciences,
  \emph{Journal of computational neuroscience}, 30, 1, 45--67

\bibitem[{\textbf{Wiskott and Sejnowski}(2002)}]{wiskott2002slow}
Wiskott, L. and Sejnowski, T.~J. (2002), Slow feature analysis: Unsupervised
  learning of invariances, \emph{Neural Computation}, 14, 4, 715--770

\end{thebibliography}

\end{document}